\documentclass[aps,prx,showpacs,floatfix,twocolumn,superscriptaddress,longbibliography,footinbib]{revtex4-2}

\usepackage{float}
\usepackage{color}
\usepackage{bm}
\usepackage{hyperref}
\usepackage{todonotes}
\usepackage{verbatim}
\usepackage{soul}
\usepackage{braket}
\usepackage{amsfonts}
\usepackage{glossaries}
\usepackage{sidecap}
\sidecaptionvpos{figure}{c} 
\usepackage{hyperref}
\usepackage{verbatim}
\hypersetup{
    colorlinks=true,
    linkcolor=blue,
    filecolor=magenta,      
    urlcolor=red,
    citecolor=blue,
}

\usepackage{tcolorbox}
\tcbuselibrary{breakable}
\usepackage{lipsum}

\newacronym{RIXS}{RIXS}{resonant inelastic x-ray scattering}
\newacronym{XAS}{XAS}{x-ray absorption spectroscopy}
\newacronym{INS}{INS}{inelastic neutron scattering}
\newacronym{IXS}{IXS}{inelastic x-ray scattering}
\newacronym{ARPES}{ARPES}{angle-resolved photo-emission}
\newacronym{EELS}{EELS}{electron energy loss spectroscopy} 
\newacronym{STS}{STS}{scanning tunneling spectroscopy}
\newacronym{SOC}{SOC}{spin-orbit coupling}
\newacronym{QSL}{QSL}{quantum spin liquid}
\newacronym{XFEL}{XFEL}{x-ray free electron laser}
\newacronym{ED}{ED}{exact diagonalization}
\newacronym{EPC}{EPC}{electron-phonon coupling}
\newacronym{CDW}{CDW}{charge-density wave}
\newacronym{SDW}{SDW}{spin-density wave}
\newacronym{SC}{SC}{superconducting}
\newacronym{KH}{KH}{Kramers-Heisenberg}
\newacronym{QMC}{QMC}{quantum Monte Carlo}
\newacronym{UCL}{UCL}{ultra-short core-hole lifetime}
\newacronym{DMRG}{DMRG}{density matrix renormalization group}
\newacronym{DMFT}{DMFT}{dynamical mean-field theory}
\newacronym{AFM}{AFM}{antiferromagnetic}
\newacronym{DCA}{DCA}{dynamical cluster approximation}
\newacronym{BCS}{BCS}{Bardeen, Cooper, and Schrieffer}
\newacronym{BZ}{BZ}{Brillouin zone}
\newacronym{1D}{1D}{one-dimensional}
\newacronym{2D}{2D}{two-dimensional}
\newacronym{3D}{3D}{three-dimensional}
\newacronym{QFI}{QFI}{quantum Fisher information}
\newacronym{trRIXS}{trRIXS}{time-resolved RIXS}
\newacronym{LCLS}{LCLS}{Linac Coherent Light Source}

\begin{document}

\title{Exploring Quantum Materials with Resonant Inelastic X-Ray Scattering}

\author{M. Mitrano}
\affiliation{Department of Physics, Harvard University, Cambridge, Massachusetts 02138, USA}

\author{S.~Johnston}
\affiliation{Department of Physics and Astronomy, The University of Tennessee, Knoxville, Tennessee 37966, USA\looseness=-1}
\affiliation{Institute for Advanced Materials and Manufacturing, The University of Tennessee, Knoxville, Tennessee 37996, USA\looseness=-1}

\author{Young-June Kim}
\affiliation{Department of Physics, University of Toronto, Toronto, ON M5S 1A7, Canada}

\author{M. P. M. Dean}\email[]{mdean@bnl.gov}
\affiliation{Condensed Matter Physics and Materials Science Department, Brookhaven National Laboratory, Upton, New York 11973, USA}

\date{\today}

\begin{abstract}
Understanding quantum materials --- solids in which quantum-mechanical interactions among constituent electrons yield a great variety of novel emergent phenomena --- is a forefront challenge in modern condensed matter physics. This goal has driven the invention and refinement of several experimental methods, which can spectroscopically determine the elementary excitations and correlation functions that determine material properties. This Perspectives article focuses on the future experimental and theoretical trends of resonant inelastic x-ray scattering (RIXS), which is a remarkably versatile and rapidly growing technique for probing different charge, lattice, spin, and orbital excitations in quantum materials. We provide a forward-looking introduction to RIXS and outline how this technique is poised to deepen our insight into the nature of quantum materials and their emergent electronic phenomena.

\end{abstract}

\maketitle


\section{Introduction}\label{sec:introduction}

\subsection{The challenges posed by quantum materials}
``Quantum materials'' is a relatively new classification intended to unify a diverse set of materials governed by many-body interactions and quantum mechanics manifested over large energy ranges and length scales \cite{Keimer2017physics, Basov2017towards, Giustino2020quantum}. While the relevant interactions in these materials are known and relatively simple, their quantum many-body nature gives rise to a host of emergent phenomena, including unconventional superconductivity~\cite{Keimer2015from}, unconventional or strange metallicity~\cite{Phillips2022stranger}, \glspl*{QSL}~\cite{Broholm2020quantum}, non-equilibrium states without equilibrium analogs~\cite{Torre2021colloquium}, and beyond.

Many quantum materials exhibit strong electron-electron interactions, which means that they often cannot be modeled accurately using textbook single-particle approximations. Instead, their correlated nature makes them prone to striking collective responses and extremely sensitive to external tuning parameters like applied fields, pressure, or doping. It is also common for different degrees of freedom to become entangled or intertwined in nontrivial ways, making it difficult to identify which interaction(s) drive a particular phenomenon. While this sensitivity represents a powerful asset for technological applications, it challenges many established concepts in condensed matter physics and hinders our predictive capabilities. For example, we still lack a comprehensive framework for many of the correlated electron liquid states that emerge from doping different ordered states. Solving this issue will require a deeper understanding of the elementary charge excitations that determine a material's dynamics, and is a key step towards determining how and why unconventional superconductivity emerges from these correlated states \cite{Keimer2015from}. More broadly, the rich physics of quantum materials poses fundamental questions about how to characterize and ultimately control their quantum states. For example, how do we definitively detect and quantify quasiparticle fractionalization and quantum entanglement in pursuit of realizing \glspl*{QSL}? How can we determine which degrees of freedom are essential to a particular state, which are helpers, and which are irrelevant? 

Beyond addressing fundamental questions about the nature of condensed matter, the field of quantum materials also strives to develop completely new states and functionalities. Advances in laser technology offer routes to hybridize materials with strong light fields to form strongly correlated electron–photon systems \cite{Bloch2022strongly}. How can we use these light-matter interactions to engineer novel transient phases? The synthesis of artificial materials and heterostructures has enabled us to observe emergent properties by exploiting proximity effects between different quantum materials in reduced dimensions. The search for new functional quantum materials has been further invigorated by the discovery of magnetic van der Waals materials, which provide platforms for new types of functional magnetic excitons. These excitations offer new routes for transducing magnetic and optical information and for realizing switchable electronic devices \cite{Bae2022exciton, Burch2018magnetism}. The path to understanding and manipulating magnetic van der Waals materials presents new challenges involving accessing the small sample volumes and buried interfaces, but this emerging area is ripe for new discoveries at the interface of applied and fundamental science.

\begin{figure*}
\includegraphics[width=1.5\columnwidth]{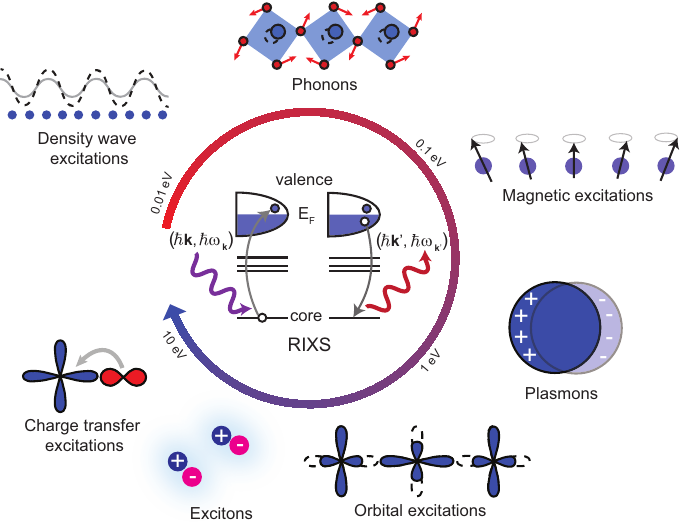} %
\caption{\textbf{The Kramers-Heisenberg process for \gls*{RIXS} and the different excitations that it can probe.} The \gls*{RIXS} process, shown in the center, involves the resonant absorption of an x-ray photon, creating an intermediate state with a core hole and a valence excitation, before the hole hole is filled via the emission of an x-ray photon. By measuring the energy and momentum change of the x-rays, one can infer the properties of the excitations created in the material. Around the outside, we illustrate the many different types of excitation that \gls*{RIXS} can probe, arranged clockwise in order of increasing energy scale, as denoted by the red-to-blue circular arrow.\glsresetall}
\label{fig_schematic}
\end{figure*}

Solving these frontier issues in quantum materials research will ultimately require us to unravel the interplay of the relevant electronic interactions along with aspects of topology, low-dimensionality, and subtle lattice effects. To this end, it has been fruitful to conceptualize different quantum materials using effective models that describe the ``essential'' aspects of a given system, namely their ground state and low-energy elementary quasiparticle and collective excitations~\cite{Pines1999elementary, Cohen2006looking}. Starting from experiments that directly measure the properties of elementary collective excitations (magnons, phonons, orbitons, etc.), and in particular their momentum dependence, one can construct highly predictive effective theories with relatively few input parameters. However, the success of this approach is critically dependent on the availability of high-quality momentum-resolved experimental probes capable of accessing the elementary excitations of spin, charge, orbital, and lattice degrees of freedom. This need has motivated the development of a host of novel experimental techniques. This Perspective shines a light on one such technique, \gls*{RIXS} (see Fig.~\ref{fig_schematic}), which has experienced rapid growth as a momentum-resolved probe of quantum materials and is endowed with unique capabilities for interrogating their collective excitations. Progress in instrumentation means that we are now at a watershed period of being able to apply \gls*{RIXS} with time and energy resolutions that match the fundamental energy scales of many quantum materials and solve key problems in this major area of condensed matter physics.

\subsection{Role of RIXS in the experimenter's toolbox}\label{sec:role_of_rixs}
An effective spectroscopic probe of quantum materials should ideally possess several attributes. Given that these solids feature multiple charge, lattice, spin, and orbital degrees of freedom, it is advantageous to be able to address all of these selectively. Furthermore, since many elementary excitations have propagating character and are thus strongly momentum-dependent, this tool should possess good momentum and energy resolution. Many questions in quantum materials hinge on measuring the properties of small samples. Therefore, it is also useful to have a probe with a suitably small focal spot size, while still being able to penetrate beneath the surface of a sample to access bulk properties. Finally, the flexibility to measure in extreme environments, such as in situ during applied electrical, magnetic, or strain fields, in a time-resolved modality after laser excitation, or within specific layers of a heterostructure, opens unique scientific opportunities. With some caveats, \gls*{RIXS} can help address all these needs. 

The central panel of Fig.~\ref{fig_schematic} illustrates the \gls*{RIXS} process. In \gls*{RIXS}, a monochromatic beam of x-rays is directed onto a material with an energy tuned to an atomic transition between a deep core state and an unoccupied valence state. The incoming photons are absorbed and excite a core electron into an unoccupied state, before the core hole is filled via the emission of another photon. By measuring the energy and momentum change of the scattered x-rays, one can extract the momentum-dependent excitation spectrum of the material. As we will explain later, the resonance process provides a key advantage of \gls*{RIXS} --- that it couples to magnetic, orbital, lattice, and charge modes as shown in Fig.~\ref{fig_schematic}. It also makes \gls*{RIXS} element-selective, which means that it can, for example, selectively study different layers in a heterostructure. 

Being an x-ray photon-in, photon-out process gives \gls*{RIXS} several advantages. The x-ray wavelength is comparable to the interatomic spacing in materials, which enables the direct measurement of the dispersion of excitations, a capability beyond the reach of infrared \cite{Basov2011electrodynamics} or Raman spectroscopy \cite{Devereaux2007inelastic}. 
The x-rays also penetrate between a few microns and about a tenth of a micron into the sample, depending on their energy, making the technique bulk sensitive.   The x-ray beams can also be focused tightly into small spots, which is needed for studying small samples. Finally, \gls*{RIXS} also benefits from pulsed high-intensity beams produced by modern \glspl*{XFEL} to perform ultrafast pump-probe experiments.

\subsection{The RIXS cross-section}\label{sec:cross_section}
Because \gls*{RIXS} involves a resonant scattering process, it is commonly described using the \gls*{KH} formalism, which is the result of treating the photon-matter interaction using second-order perturbation theory \footnote{For a detailed derivation of this expression from the light-matter interaction, we refer the reader to the review by Ament {\it et al}.~\cite{Ament2011resonant}}. 
Denoting the momentum, energy, and polarization of the incoming (outgoing) x-rays 
as $\hbar \bm{k}$, $\hbar\omega_{\bm{k}}$, and $\hat{\epsilon}$ ($\hbar \bm{k}^\prime$, $\hbar\omega_{\bm{k}^\prime}$, and $\hat{\epsilon}^\prime$), respectively, the intensity for \gls*{RIXS} is proportional to 
\begin{equation}\label{eq:KH}
I \propto \sum_f |M_{fi}|^2 \delta(E_f-E_i-\hbar\omega). 
\end{equation}
In this process we define, photon energy loss $\hbar\omega_{\bm{q}} = \hbar(\omega_{\bm{k}}-\omega_{\bm{k}^\prime})$ and scattering vector $\hbar \bm{q} = \hbar(\bm{k}^\prime - \bm{k})$. $M_{fi}$ is the matrix element from the system's initial state $i$, with energy $E_i$, to its final state $f$, with energy $E_f$, via an intermediate state $n$ with a core hole. In the \gls*{KH} approach, the matrix element is given by 
\begin{equation}\label{eq:KH_ME}
M_{fi} = \sum_n \frac{\bra{f} {\cal D}^\dagger_{\bm{k}^\prime,\hat{\epsilon}^\prime}\ket{n}\bra{n} {\cal D}^{\phantom\dagger}_{\bm{k},\hat{\epsilon}}\ket{i}}{E_n - E_i - \hbar\omega_{\bm{k}} - \mathrm{i}\Gamma_n/2},
\end{equation}
where $\Gamma_n/2$ is the inverse core-hole lifetime in units of energy, $E_n$ is the energy of the intermediate state, and $\cal{D}^{\phantom\dagger}_{{\bf k},\hat{\epsilon}}$ ($\cal{D}^\dagger_{{\bf k}^\prime,\hat{\epsilon}^\prime}$) is the absorption (emission) operator for the appropriate edge. Since the x-ray wavelength tends to be larger than the extent of the atomic orbitals, these operators are often evaluated using the dipole approximation.  

While Sec.~\ref{sec:Theory} discusses the theoretical treatment of the \gls*{RIXS} cross-section, Eqs.~\eqref{eq:KH} and \eqref{eq:KH_ME} already provide several insights. The scattering process involves a quantum mechanical superposition of all possible intermediate states $\ket{n}$ distributed throughout the lattice, which is why the collective, momentum-dependent excitations of a material can be measured. In this sense, \gls*{RIXS} is related to the more established technique of \gls*{INS}, which is a powerful probe of magnetic excitations and phonons \cite{Squires2012introduction} (see also Box 1).

A common approach to treating experimental scattering cross-sections is to map them onto multiparticle response functions (which only depend on the material) multiplied by prefactors accounting for the probe's attributes. \Gls*{ARPES} reflects the single-particle spectral function multiplied by a factor related to the dipole interaction of incident light with the material \cite{Damascelli2003angle, Sobota2021angle}. Magnetic \gls*{INS}, on the other hand, involves a two-spin correlation function multiplied by a simple analytical prefactor that depends on the neutrons \cite{Squires2012introduction}. This description thus streamlines theoretical calculations and enables them to be directly compared with experimental observations. \gls*{RIXS} is different and, as we discuss in Sec.~\ref{sec:Theory}, there is no completely general way to make such a simplification for this probe. Often, elementary excitations are identifiable from basic considerations, such as the nature of the material under study or the energy scale and dispersion of the excitation itself, but this is not always the case.
 
The \gls*{RIXS} cross-section complexity, however, provides opportunities to extract additional information. The visibility of certain excitations depends on the type of core hole resonance such that specific excitations can be isolated through the judicious selection of a particular resonance energy. For example, choosing an intermediate state with a $2p$ core hole, which features strong \gls*{SOC}, allows the exchange of the orbital angular momentum of the photon with the spin angular momentum in the valence band to directly create spin-flip excitations \cite{Braicovich2010magnetic}. At a $1s$ core hole resonance, on the other hand, single spin-flip excitations are forbidden \cite{Hill2008observation}. Specific choices of the incoming and scattered x-ray polarization can selectively excite spin or orbital excitations according to the x-ray absorption and emission operators \cite{Ament2009theoretical}. The choice of the resonance can also emphasize either direct ``operator'' \gls*{RIXS}, where the excitations arise from the dipole operators, or indirect  ``shake-up'' RIXS, where excitations are created largely by interactions between the core hole and the valence electrons. Most excitingly, this richness allows \gls*{RIXS} to access excitations that cannot or have not been measured in any other way. Examples of these include dispersive orbital excitations called ``orbitons'' \cite{Schlappa2012spin}, quintuple magnons \cite{Li2023single, Elnaggar2023magnetic}, excitations involving coupled \glspl*{CDW} and phonons \cite{Chaix2017dispersive}, and fractional excitations in Kitaev materials \cite{Halasz2016resonant}.

\subsection{Outline}
The goal of this Perspective is to shine a spotlight on key topics for future \gls*{RIXS} studies of quantum materials. In light of this focus, we refer any reader interested in applications to liquids, gases, or solid-state chemistry to recent reviews on these topics~\cite{Lundberg2019resonant, Gelmukhanov2021dynamics}. Section~\ref{sec:historical} also provides introductory historical context on the development of \gls*{RIXS} as a probe of quantum materials and the boxed section \emph{Comparison between \gls*{RIXS} and other scattering techniques} provides some context comparing \gls*{RIXS} to other techniques. A reader who is already familiar with this material can safely skip these sections. 

The remainder of this Perspective focuses on a selection of topics that we consider to be particularly important for the field of quantum materials. Section~\ref{sec:strange} explores how \gls*{RIXS} measurements of charge excitations will contribute to the paradigmatic challenge of understanding the anomalous properties of strongly correlated metals, including so-called ``strange metals,'' states characterized by poorly understood electrical conduction and symmetry-breaking properties. Section~\ref{sec:QSL} addresses  \gls*{QSL} materials and outlines \gls*{RIXS}'s prospects for detecting their novel excitations, which will provide new insights into their unique properties like carrier fractionalization and many-body entanglement. The advent of \glspl*{XFEL} has opened remarkable new prospects for understanding and controlling dynamical phenomena arising from light-matter interactions. In light of this, Sec.~\ref{sec:non-equilibrium} examines future opportunities with \gls*{trRIXS}, which is uniquely positioned to study the collective excitations of these driven systems and prove the existence of novel types of non-equilibrium states. Section~\ref{sec:functional_materials} examines what insights \gls*{RIXS} can bring to the study of functional quantum materials. Here, we place a particular focus on novel excitons (collective excitations of bound electron-hole pairs) and their dynamics, which offer potential for fundamentally new forms of magneto-optical coupling. Finally, Sec.~\ref{sec:Theory} covers key concepts and challenges in the theory of \gls*{RIXS}, which will be crucial to advancing the field.

\begin{figure*}
\justifying
\begin{tcolorbox}[width=\textwidth, colback=gray!5!white,colframe=black!75!black,title=BOX 1: Comparison between RIXS and other scattering techniques, parbox=false]
\textbf{\gls*{RIXS}} is just one member of a family of scattering techniques suited to measuring the collective excitations of quantum materials, distinct from single-particle probes such as \gls*{ARPES}. In some ways, its closest cousin is \textbf{\gls*{IXS}}, a non-resonant technique. Away from core-level resonances, x-rays interact with the sample primarily through the Thomson cross-section, where the x-ray's electric field interacts with all the valence and core electrons in the material. Since most electrons are tightly bound to nuclei deep within the atoms, \gls*{IXS} is a highly effective probe of phonons \cite{Baron2020introduction}. It also couples to plasmons, but the cross-section is a function of the ratio of valence to core electrons making such measurements practical only for materials made up of light elements \cite{Schuelke2007electron}. \gls*{IXS} also has negligible coupling to many of the other excitations shown in Fig.~\ref{fig_schematic}, including magnetic excitations. The x-ray energy used in \gls*{IXS} can be chosen to optimize the energy resolution of the experiment, rather than being fixed by the core hole resonance, and \gls*{IXS} setups can achieve sub-meV energy resolution using spectrometers based on back-scattering crystal optics \cite{Baron2020introduction}.
\vspace{0.5em}

\textbf{\Gls*{EELS}} is another analogous technique involving the inelastic scattering of electrons rather than photons. Because electrons are charged particles, they interact very strongly with the electronic clouds of a solid and are naturally sensitive to a larger variety of valence excitations \cite{Ibach2013electron, Abbamonte2024collective}. \gls*{EELS} experiments are performed both in transmission and reflection geometries. Transmission \gls*{EELS} (T-EELS) measurements have a rather simple cross-section \cite{Pines1989theory, Pines1999elementary} and rely on keV-energy electrons to penetrate into the bulk of nm-thick samples. Reflection \gls*{EELS} (R-EELS) experiments typically involve lower energy ($\sim 10-100$ eV) electrons, which scatter off the surface of solids. Owing to the lower energy, R-\gls*{EELS} experiments can straightforwardly achieve meV energy resolution. However, the interaction with the surface can complicate the interpretation of the cross-section of bulk quantum materials and limit access to out-of-plane momenta \cite{Vig2017measurement, Abbamonte2024collective}.
\vspace{0.5em}

Another important technique is \textbf{\acrfull*{INS}}. Neutrons scatter from both the atomic nuclei and valence electron magnetic moments, which interact with the neutron's magnetic moment. As such, \gls*{INS} is well-suited to measuring phonons and magnetic excitations. The weak interactions between thermal neutrons and solids allows one to connect the signal to two-spin correlation functions \cite{Squires2012introduction}, which facilitates the straightforward quantitative interpretation of data. However, the same weak interactions also mean that large samples are needed to obtain adequate signals. \gls*{INS} is especially well suited to studying very low energy excitations and can be routinely applied with sub-meV or even µeV resolution.
\vspace{0.5em}

When comparing \gls*{RIXS} with these other scattering methods, it is important to note that the \textbf{achievable performance} in \gls*{RIXS} experiments is strongly dependent on the incident x-ray energy, including whether the energy falls within the soft, tender, or hard x-ray regimes. Most high-resolution \gls*{RIXS} experiments on quantum materials today are performed with energy resolution of $20$--$40$~meV, which is well matched with charge or orbital excitations and magnetic excitations in materials with large exchange energies. The x-ray energy also determines the accessible momentum range. Typically, intermediate or hard x-ray \gls*{RIXS} can cover the whole (and multiple) \gls*{BZ} and soft x-ray \gls*{RIXS} often only provides partial coverage. 
\vspace{0.5em}

These techniques also offer different possibilities for \textbf{extreme sample environments}. The penetrating power of neutrons makes \gls*{INS} relatively easy to implement at high magnetic field and dilution fridge temperatures, while small beam size makes hard x-ray \gls*{RIXS} compatible with studying materials under high pressure using a diamond anvil cell. The vacuum requirements, the limited availability of window materials, and the proximity of complex goniometers and/or spectrometer components makes extreme sample environments more difficult (but far from impossible) to implement for soft x-ray \gls*{RIXS}. \gls*{RIXS} is typically performed in the 10-300~K range and under either zero or modest magnetic fields of a fraction of a Tesla. Improving these aspects of the sample environment is particularly important for the study of \glspl*{QSL}. \gls*{EELS} is usually performed in vacuum and over a similar temperature range, but cannot be combined with high pressures (due to the electron penetration depth) and applied magnetic fields (since these would dramatically deflect the electrons).
\end{tcolorbox}

\end{figure*}

\section{A very brief history of RIXS}\label{sec:historical}
As with any subject, the history of \gls*{RIXS} is helpful in understanding future trends in this area of research. \gls*{RIXS} is a challenging experimental technique at the forefront of x-ray science. Being a second-order process, it has a weak cross-section, necessitating a bright x-ray source and an efficient detection scheme. The energy scale of x-ray photons is also far larger than the typical excitation energies in quantum materials, which means that high resolving powers (defined as the incident photon energy divided by the energy resolution) are required. For these reasons, the development of \gls*{RIXS} is in large part dictated by the pace of development of x-ray sources and spectrometers and can be thought of as a synergistic evolution of technological and scientific breakthroughs. In judging this progress, it is useful to consider that a typical order-of-magnitude temperature scale for phase transitions in many quantum materials is $T\sim100$~K, which corresponds to an energy scale of $k_\text{B} T \sim 10$~meV and a timescale of $\hbar / (k_\text{B} T) \sim 100$~fs (where $k_\text{B}$ is Boltzmann's constant). A key reason why \gls*{RIXS} is increasingly influential in the field of quantum materials is that its resolution is now approaching this important threshold. However, energy resolution remains a limiting factor in many types of experiments.

The way that \gls*{RIXS} instrumentation energy-resolves x-rays falls into two general types, depending on the energy of the x-ray resonance they target. This means that \gls*{RIXS} experimentalists must carefully consider the resonant energy of the target core hole transition. For so-called ``hard'' x-ray energies above about 5~keV, x-ray wavelengths ($\lesssim 3$~\AA{}) are comparable to the lattice spacing of nearly-perfect crystals such as silicon, and spectrometers in this energy range are typically based on crystal analyzers. Below about 2~keV, x-ray wavelengths are too long for this to work effectively and soft x-ray spectrometers are consequently based around gratings \cite{Dvorak2016towards, Brookes2018beamline, Zhou2022I21, Singh2021development}. Efforts to perform experiments in the more difficult energy range of about 2-5~keV, in what is sometimes called the tender x-ray regime, are opening compelling new research opportunities \cite{Gretarsson2020IRIXS, Bertinshaw2021IRIXS,Suzuki2023distinct}.

Some of the first \gls*{RIXS} research dates to the 1970s, when Eisenberger, Platzman, and collaborators demonstrated the resonant enhancement of inelastic x-ray scattering from copper metal \cite{Eisenberger1976xray, Eisenberger1976resonant}. However, the energy resolution at the time (0.8~eV) made it quite difficult to infer new details about copper's electronic properties. In the mid-1990s, \gls*{RIXS} started becoming an incisive probe of quantum materials, fueled by the marriage of improved insertion device x-ray sources and dedicated spectrometers alongside the realization that it can directly measure the energy of electronic transitions. Initial studies focused on band insulators and semiconductors \cite{Rubensson1990excitation, Ma1992soft, Kotani2001resonant, Schuelke2007electron}. Not long after, the exploration of materials such as NiO \cite{Kao1996xray, Braicovich1997xray, Groot1998local} and CeO$_2$ \cite{Butorin1996resonant} initiated a new domain of \gls*{RIXS} research focused on materials with strong electronic correlations.

Buoyed by interest in strongly correlated quantum materials and cuprate high-temperature superconductors in particular \cite{Dean2015insights}, as well as developments in theory, \gls*{RIXS} has evolved from a niche to a flourishing field. A series of works in the late 1990s and early 2000s established its unique capabilities to resolve the momentum dependence of insulating ``Mott'' gap excitons \cite{Hill1998resonant, Abbamonte1999resonant, Hasan2000electronic, Kim2002resonant} and the crystal field of cuprates \cite{Ghiringhelli2004low}. Driven by a two-orders-of-magnitude improvement in both resolution and count rates \cite{Ament2011resonant, Dvorak2016towards, Brookes2018beamline, Zhou2022I21, Singh2021development, Shvydko2013MERIX}, the next decade saw breakthrough experimental demonstrations that \gls*{RIXS} can measure magnetic excitations \cite{Hill2008observation, Schlappa2009collective, Ghiringhelli2009, Braicovich2010magnetic}. These investigations focused first on undoped and underdoped cuprates \cite{Braicovich2010magnetic, Guarise2010measurement} but quickly spread to optimally-doped, overdoped, and electron-doped compositions \cite{LeTacon2011intense, Dean2013high, Dean2013persistence, Lee2014asymmetry, Ishii2014high}, as well as to other magnetic materials \cite{Zhou2013persistent, Fabbris2017doping, Pelliciari2021tuning}. Here, \gls*{RIXS} was key in clarifying the nature of magnetic correlations and their relationship with high-temperature superconductivity \cite{Dean2015insights}. A similar explosion of activity occurred around iridates \cite{Kim2012magnetic, Kim2012large, Yin2013ferromagnetic, Gretarsson2013crystal, Kim2014excitonic, Chun2015direct}. \gls*{RIXS} was key to establishing how strong \gls*{SOC} in these materials can realize novel types of correlated insulating and magnetically frustrated states \cite{Kim2008novel, Jackeli2009Mott}. Today, measuring the spin excitations and magnetic interactions of a material remains one of \gls*{RIXS}'s main applications within quantum materials research.  

Major results that go beyond magnetic excitations included the identification of dispersive orbital excitations or orbitons \cite{Schlappa2012spin}, which arise from a special quantum effect of ``spin–orbital separation'' in one dimension. \gls*{RIXS} was also central to establishing the ubiquitous presence of \glspl*{CDW} in cuprates \cite{Ghiringhelli2012long, Tabis2014charge, Miao2017high, Chaix2017dispersive, Peng2018re, Arpaia2019dynamical, Lin2020strongly, Yu2020unusual, Tam2022chargecuprates}, supporting the contention that this symmetry breaking is a fundamental aspect of the physics of these materials. Demonstrations that \gls*{RIXS} can probe electron-phonon coupling \cite{Ament2011determining, Lee2013role, Rossi2019experimental}, spin excitations in atomically thin layers within heterostructures \cite{Dean2012spin} or on substrates \cite{Pelliciari2021evolution}, and ultrafast transient states \cite{Dean2016ultrafast, Mitrano2019ultrafast} have further established some of the unique capabilities and potential of the technique. 

\section{Strongly correlated metals}\label{sec:strange}
\begin{SCfigure*}
\includegraphics[width=0.6\textwidth]{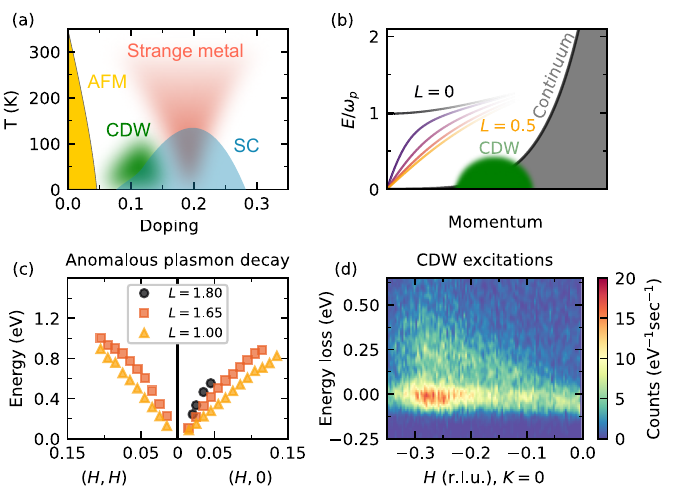}
\caption{\textbf{The charge excitation spectrum of correlated metals holds the key to understanding their anomalous electronic transport and symmetry breaking properties.} (a) Electronic phase diagram of the cuprates showing the emergence of the unconventional or strange metal phenomenology from neighboring \gls*{CDW}, superconducting, and magnetic phases. (b) Conceptual picture of the charge excitations of cuprates, which reflect the properties of these unusual metalic states. At low momentum, cuprates feature plasmons, but at higher momentum, these decay to form a continuum in poorly understood ways. Cuprates also universally exhibit \gls*{CDW} symmetry breaking and correlations. (c) Plasmon dispersion and decay in Nd$_{2-x}$Ce$_x$CuO$_4$ \cite{Hepting2018three}. (d) Low-energy \gls*{CDW} excitations in La$_{2-x}$Ba$_x$CuO$_4$ above the \gls*{CDW} transition \cite{Miao2017high}. Understanding these excitations can be key to comprehending unconventional metallicity.}
\label{fig_strange_metals}
\end{SCfigure*}

Anomalous metallic behavior beyond the Fermi liquid paradigm is ubiquitous in quantum materials \cite{Paschen2021quantum}. For example, many compounds exhibit unconventional electron-electron scattering that causes their resistivity to scale linearly with temperature over a very large range, unlike the $T^2$ dependence predicted by Fermi liquid theory. This phenomenology is often called ``strange'' metallicity \cite{Phillips2022stranger, Takagi1992systematic}. Some of these materials even exhibit resistivity exceeding the Mott-Ioffe-Regel limit, where the mean free path of electrons is shorter than the interatomic spacing, which is often termed ``bad'' metal behavior~\cite{Hussey2004universality}). 

The poster child for this physics is the cuprates, where strange metal behavior emerges in the vicinity of other types of novel electronic behavior, including \gls*{CDW} symmetry breaking and unconventional superconductivity [see Fig.~\ref{fig_strange_metals}(a)]. Moreover, these types of phenomena also appear in a host of different strongly correlated materials, including ruthenates \cite{Hussey1998normal}, iron pnictides \cite{Analytis2014transport}, heavy fermion materials \cite{Paschen2021quantum}, and even magic-angle twisted bilayer graphene \cite{Cao2020strange}. 
Given the universality of this unconventional metal phenomenology, many researchers believe understanding it will require fundamentally new physical concepts or mathematical machinery \cite{Keimer2015from}. To date,  \gls*{RIXS} has played an essential role in deepening our understanding of correlated quantum materials, with the cuprates often acting as a touchstone in this endeavor. Looking toward the future, it will continue to provide momentum- and energy-resolved measurements of the collective charge excitations across many of these systems, which will be crucial for formulating and ultimately validating new theoratical frameworks for correlated metals.

\subsection{Charge excitations}
At a fundamental level, unconventional metal phenomenology suggests that the collective charge excitations in cuprates are \textit{radically different} from those predicted in the standard Fermi liquid theory \cite{Phillips2022stranger}. Understanding these excitations may hold the key to solving the strange metal problem \cite{Patel2023universal} and in formulating theories to explain other novel phenomena observed in unconventional metals \cite{Varma2020colloquium}.

In the long-wavelength ($\bm{q}\rightarrow0$) limit, metals typically exhibit well-defined electronic excitations called plasmons. These are collective oscillations in the electron density that primarily reflect the compressibility of the electron system. Within a Fermi-liquid picture, the plasmon is expected to disperse as a function of momentum, before it intersects with a continuum of particle-hole excitations, whereupon it decays. 
Optics experiments have shown that the low-$\bm{q}$ charge excitation spectra of cuprates indeed feature plasmons \cite{Bozovic1990plasmons,Levallois2016temperature}. \gls*{RIXS} experiments have recently measured their low-$\bm{q}$ dispersion in hole- and electron-doped cuprates as a function of both in-plane and out-of-plane $\bm{q}$, as plotted in Fig.~\ref{fig_strange_metals}(b) and ~\ref{fig_strange_metals}(c) \cite{Hepting2018three, Lin2020doping, Nag2020detection, Singh2022acoustic, Hepting2022gapped, Hepting2023evolution}. These plasmons, however, are broad and decay anomalously at relatively low $\bm{q}$ before they intersect the 
 particle hole continuum \cite{Mitrano2018anomalous,Husain2019crossover, Nag2020detection}. The reasons for this remain poorly understood and are not accurately predicted by the established models.

Determining the detailed form of the high-momentum charge continuum in the cuprates and how it relates to different theories for unconventional metals and the strange metal phenomenology is an area where \gls*{RIXS} is likely to make decisive contributions. An important challenge is the need to isolate the charge response from the spin response and the lattice, which may be addressed by approaches such as x-ray polarimetry or some of the theoretical advances that we anticipate in Sec.~\ref{sec:Theory}. \Gls*{EELS} is also likely to bring an important complementary perspective to these questions, as it has better energy resolution compared to \gls*{RIXS} and a simpler cross-section \cite{Vig2017measurement}. However, its surface sensitivity presents additional challenges and limits its access to the out-of-plane plasmon dispersion.

An especially topical route to understanding cuprates is to contrast them with non-copper-containing materials with closely related properties, which will help identify the key ingredients behind their novel phenomenology. The past few years have seen the discovery of superconductivity and unconventional metallicity in nickelates, including those formed from square planar nickelate oxide layers with a low electronic valence of $d^{9-\delta}$ \cite{Li2019superconductivity, Li2020superconducting, Pan2022superconductivity} and also those built from bilayers of nickel-oxygen octahedra \cite{Sun2023signatures, Zhu2024superconductivity}. Synthesizing these materials as large single crystals and preparing pristine surfaces has proven quite challenging, so \gls*{RIXS} has had an out-sized role in characterizing their electronic properties including the nature of the electronic correlations and the role of oxygen orbitals \cite{Hepting2020electronic, Lin2021strong, Lu2021magnetic, Shen2022role}. Although theoretical predictions for plasmons in these nickelates are available \cite{Zinni2023low}, to date there are no published experimental measurements. Looking toward the future, fully characterizing the collective excitation spectrum of nickelates and comparing their phenomenology with that of cuprates will constitute an important contribution towards the solution of the strange metal problem. Another element of this endeavor will be clarifying the currently controversial presence and nature of electronic symmetry breaking in the same materials, which is discussed next \cite{Rossi2022broken, Tam2022chargenickelates, Krieger2022charge, Shen2023electronic, Parzyck2023absence}.

\subsection{Electronic symmetry breaking}
Several decades of study of cuprates have established that charge and spin symmetry breaking are ubiquitous features of these materials \cite{Comin2016resonant, Frano2020charge}. Different types of charge and spin symmetry breaking are also common features of other strongly correlated metals such as nickelates, pnictides, and heavy fermion materials \cite{Fradkin2015colloquium}. Modern numerical methods predict that charge order is also a feature of simple effective models such as the Hubbard model, and support the idea that it can arise from strong electronic correlations \cite{Zheng2017stripe, huang2018stripe, Qin2020absence, Mai2022intertwined}. From a physical point of view, doped carriers can break fewer magnetic bonds if they cluster together, leading to an effective attraction between holes. This effective interaction is countered by Coulomb repulsion and hopping, which both tend to distribute carriers evenly through the crystal to minimize the total energy. The fact that these competing interactions act on different length scales implies that the overall minimum energy solution of this complicated situation may involve a modulated clustering: a \gls*{CDW}. Here, the term \gls*{CDW} correlation refers to the general \gls*{CDW} phenomenology that includes static long- and short-range order, as well as dynamic \gls*{CDW} fluctuations.

\gls*{CDW} correlations in cuprates interact strongly with other electronic phases. Although static charge and spin order appear to suppress superconductivity, some researchers have suggested that dynamic fluctuations may act to promote it \cite{Emery1997spin, Kivelson1998electronic}. Scattering from \gls*{CDW} fluctuations may have a role in the anomalous resistivity of strange metals \cite{Castellani1995singular}. \gls*{RIXS} has exceptional sensitivity to weak \gls*{CDW} correlations, as it combines the resonant enhancement of valence electron modulations, while using a spectrometer to reject the strong x-ray fluorescence that limits the sensitivity of traditional resonant soft x-ray scattering experiments \cite{Ghiringhelli2012long, Fink2013resonant}. Recently, \gls*{RIXS} has shown that diffuse \gls*{CDW} correlations extend to much higher temperatures and dopings than previously thought \cite{Miao2017high, Peng2018re, Arpaia2019dynamical, Miao2019formation, Lin2020strongly, Tam2022chargecuprates, Li2023prevailling}. Figure~\ref{fig_strange_metals}(d) plots a \gls*{RIXS} measurement of La$_{2-x}$Ba$_x$CuO$_4$ showing a broad feature centered around $H=-0.27$~r.l.u.\ and modification of the continuum at finite energies \cite{Miao2017high}. 

While the widespread presence of \gls*{CDW} correlations is well established, the nature and energy of the low-energy excitations associated with the \gls*{CDW} $\bm{q}$-vectors are not. Solving this should help definitively unravel the interactions underlying \gls*{CDW} formation. \gls*{RIXS} holds promise to disentangle these effects by examining the interactions between charge and spin \cite{Miao2017high} or detailed studies of phonons as discussed in Sec.~\ref{sec:th_EPC} \cite{Chaix2017dispersive, Peng2020enhanced, Lee2021spectroscopic}. This may help clarify how important structural distortions are in the formation of \glspl*{CDW}. Studies under applied strain can also be very insightful, especially if a material is on the verge of long-range \gls*{CDW} order \cite{Kim2018uniaxial, Kim2021charge}.

Another major question is whether the cuprate phase diagram hosts a possible \gls*{CDW} quantum critical point or \gls*{CDW} quantum criticality in general. Near a quantum critical point, we expect a divergence of the fluctuations associated with the ordering, which can mediate unconventional forms of superconductivity \cite{Monthoux2007superconductivity}. Thermodynamic and transport probes indeed provide evidence of quantum criticality, but the location of these effects in the doping phase diagram suggests that quantum criticality might be more related to the pseudogap than to a \gls*{CDW} \cite{Cooper2009anomalous, Michon2019thermodynamic}. \gls*{RIXS} studies of how \gls*{CDW} excitations scale with temperature nonetheless suggest that quantum critical effects are at play in the cuprates \cite{Lee2021spectroscopic, Huang2021quantum, Arpaia2023signature}. Determining the role of quantum criticality in the cuprates will require the identification of universal effects that are present in all superconducting cuprate families. Given that the characteristic temperature of these effects is about 100~K, it will be important to obtain a better energy resolution in the few meV energy range rather than in the few 10s of meV range. Such resolutions are being targeted in \gls*{RIXS} instruments that are currently under construction \cite{Miyawaki2022design}.

\section{Quantum spin liquids}\label{sec:QSL}  
\begin{figure*}[ht]
\includegraphics[width=0.98\textwidth]{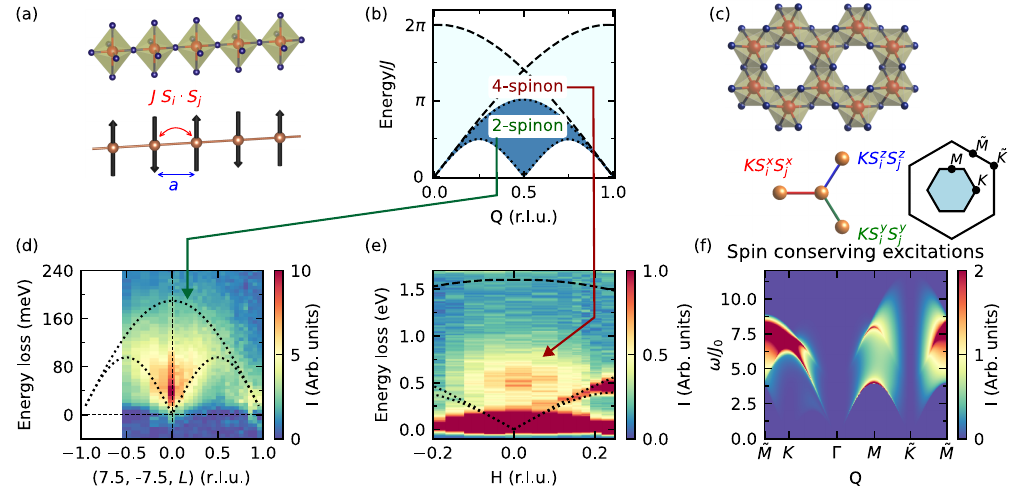}
\caption{\textbf{\gls*{RIXS} can access novel types of higher-order correlations functions that can help us identify \glspl*{QSL}.} (a) \Gls*{1D} Heisenberg spin chain. The \gls*{AFM} Heisenberg interaction $J>0$ acts equally on all three spin components. (b) Schematic diagram showing the phase space of fractionalized magnetic excitations in a spin chain. The darker shaded region is the 2-spinon continuum, and all shaded regions (both light and dark blue) represent the 4-spinon continuum \cite{Caux2006four}. (c) Kitaev model where spins connected by an $x$-bond interact via their $x$-components and equivalent for the $y$- and $z$-bonds. (d) Experimental Ir $L_3$-edge \gls*{RIXS} data of the magnetic excitations along the chain direction in Ba$_4$Ir$_3$O$_{10}$ from Ref.~\cite{Shen2022emergence}. The dotted lines are the 2-spinon continuum boundary shown in panel (b). (e) Experimental \gls*{RIXS} spectra of a spin chain compound Sr$_2$CuO$_3$ obtained at the oxygen $K$-edge \cite{Schlappa2018probing}. The dotted and dashed lines indicate the boundaries of the two- and four-spinon continua, respectively, as shown in panel (b). (f) Theoretical \gls*{RIXS} intensity of the Kitaev honeycomb model in the spin-conserving channel showing a strongly structured spectrum consistent with spin fractionalization \cite{Halasz2016resonant}. The symmetry notation for the \gls*{BZ} is shown above the plot.}
\label{fig_qm}
\end{figure*}

\glspl*{QSL} are a fundamental topic in magnetism. These are magnetic materials in which quantum fluctuations suppress the tendency to form classical long-range magnetic order, even in the limit of zero temperature \cite{Zhou2017quantum, Savary2017Quantum, Broholm2020quantum}. To understand \glspl*{QSL} it is useful to consider a \gls*{1D} spin-$\frac{1}{2}$ Heisenberg antiferromagnet model, as shown in Fig.~\ref{fig_qm}(a). The model's Hamiltonian is well understood theoretically and captures the essential physics of several real materials like Sr$_2$CuO$_3$ \cite{Bethe1931theory, Cloizeaux1962spin, Walters2009effect}. Instead of a N\'{e}el-ordering and magnon excitations, as expected from a classical picture, the system's ground state is an entangled quantum many-body state with fractional quasiparticle excitations called spinons. Due to the fact that these fractionalized quasiparticles carry spin-$\frac{1}{2}$, the magnetic excitation spectrum of a material hosting spinons involves exciting even numbers of spinons, creating broad continua, as plotted in Fig.~\ref{fig_qm}(b).

The key goal of \gls*{QSL} research is to definitively realize quantum entangled states with fractional quasiparticles in \gls*{2D} or \gls*{3D} materials. A well-known way to do this is through geometric frustration, which is found when no ordered state can simultaneously satisfy all the interactions between different spin pairs. Another type of frustration that has drawn much attention recently is exchange frustration caused by bond-directional interactions called Kitaev interactions. Kitaev's honeycomb model shown in  Fig.~\ref{fig_qm}(c) is exactly solvable and features a \gls*{QSL} ground state with fractional excitations called Majorana fermions, a potentially revolutionary component for topological quantum computation \cite{Kitaev2006anyons, Nussinov2015compass}. Several iridates with honeycomb or honeycomb-like lattices, as well as $\alpha$-RuCl$_3$, have been suggested as candidates for Kitaev \glspl*{QSL} \cite{Winter2017models, Trebst2022kitaev, Motome2020hunting, Takagi2019concept, Kim2022alpha}. Many \gls*{INS} and \gls*{RIXS} experiments have been performed on these materials, most of which are \emph{consistent} with what is expected for a \gls*{QSL} \cite{Banerjee2016proximate, Banerjee2017neutron, Gretarsson2013crystal, Chun2015direct, Torre2021enhanced, Lebert2020resonant, Suzuki2021proximate}. The key open challenge is to \emph{definitively} identify \glspl*{QSL}. This can be difficult because these states tend to exhibit broad and somewhat featureless continua, which, in isolation, can be hard to distinguish from classical spin glass states or the effects of disorder \cite{Shen2016Evidence, Paddison2017continuous, Ma2018spinglass}. In the following, we outline two ways to address this using RIXS: the first is using special features of the \gls*{RIXS} cross-section to identify spectroscopic signatures of electron fractionalization, and the second is to combine \gls*{RIXS} with concepts from quantum information to directly prove the presence of quantum entanglement.

\subsection{Spectroscopic signatures of QSLs}
While \gls*{INS} is an indispensable tool for studying quantum magnetism, \gls*{RIXS} can provide additional information about a material's magnetic excitations. This capability arises from the \gls*{RIXS} cross-section, which allows for excitation pathways beyond the conventional dipolar spin flips generated by \gls*{INS}. The cases below illustrate how the richness of the \gls*{RIXS} cross-section will offer routes to more definitively probe \glspl*{QSL}. 

As a first example, it is useful to compare the \gls*{RIXS} spectra of spinon-hosting materials taken at the $L$- and $K$-edges shown in Fig.~\ref{fig_qm}(d, e), respectively. At the $L$-edge, the spectrum is dominated by a 2-spinon continuum yielding similar information to that obtained by \gls*{INS}. The $K$-edge, on the other hand, features a strong indirect \gls*{RIXS} process, which produces 4-spinon excitations that are well separated from the 2-spinon continuum and that are not observed using \gls*{INS}. These higher-order correlations have a distinct structure that can be exploited to better identify \glspl*{QSL}. Proof of principle of this approach has been achieved by studying \gls*{1D} chains \cite{Schlappa2018probing, Kumar2021Unraveling}, but to date, this has not yet been applied to \gls*{2D} or \gls*{3D} \gls*{QSL} candidates. In the same vein, the quadrupole transition of \gls*{RIXS}, which gives rise to $\Delta S_z =2$ excitations via double spin-flips in the same atom, offers additional routes to more deeply interrogate \gls*{QSL} candidates with spin $>1/2$. Such $\Delta S_z =2$-type excitations were theoretically proposed and experimentally observed in quantum magnets 
\cite{Haverkort2010theory,Ament2011resonant, Nag2020many, Nag2022quadrupolar}. Recent studies reported that it is even possible to create multiple spin-flips corresponding to $\Delta S_z =3,4,5$ \cite{Li2023single, Elnaggar2023magnetic}. 

Another important distinction between \gls*{INS} and \gls*{RIXS} is that the latter can couple to an electron's charge and orbital degrees of freedom, in addition to its spin degree of freedom. This becomes particularly important for probing fractional particles such as Majorana fermions. To identify fractionalized excitations uniquely, additional information encoded in the incoming and outgoing photon polarization (a so-called polarimetry analysis) of the \gls*{RIXS} spectrum is valuable \cite{Halasz2016resonant,Halasz2019observing, Halasz2017probing, Natori2020dynamics, Natori2017dynamics, Ko2010raman, Rousochatzakis2019quantum, Fu2021dynamic}. Theoretical predictions of this effect are shown in Fig.~\ref{fig_qm}(f), which plots \gls*{RIXS} calculations for a Kitaev spin liquid in the ``spin-conserving'' polarization channel \cite{Halasz2016resonant}. Strongly momentum-dependent spectral structure is seen coming from fractionalized quasiparitcles. This is distinct from the featureless spectrum in the ``non-spin-conserving'' channel, corresponding to the usual spin structure factor probed by \gls*{INS}. The power of polarimetry analysis was demonstrated in a study of a two-dimensional copper oxide compound, where the emergence of spinons from the decay of high-energy magnons was identified by analyzing the outgoing photon polarization \cite{Martinelli2022fractional}. One challenge in this area is that polarimetry setups have limited efficiency, so it can be difficult to perform the polarization analysis while maintaining the high energy resolution and signal strength required to observe the desired features \cite{Braicovich2014simultaneous}. Advances in x-ray multilayer optics and the inclusion of collimating mirrors in modern spectrometers are expected to improve the efficiency of polarimetry measurements in the coming years \cite{Brookes2018beamline, Kim2016collimating}. Another challenge is that the energy resolution of \gls*{RIXS} is currently significantly inferior to \gls*{INS} (see Box 1) so these studies are likely to focus on transition oxide materials due to their comparatively large energy scales. 

\subsection{Witnessing entanglement}\label{sec:witnessing}
A major opportunity for future research lies in the marriage of scattering probes and quantum information methods. The spectral response of materials in the thermodynamic limit can be rigorously connected to the expectation values of operators called ``entanglement witnesses,'' which certify the presence of quantum entanglement and thus have the potential to more directly and definitively test whether materials are \glspl*{QSL} \cite{Brukner2006crucial, Coffman2000distributed, Hyllus2012fisher,Amico2006divergence, Amico2008entanglement, Hauke2016measuring}. Recent \gls*{INS} studies have demonstrated the power of this method in low-dimensional spin systems \cite{Mathew2020experimental, Scheie2021witnessing, Laurell2021quantifying, Scheie2024witnessing, Menon2023multipartite}. These studies are based on a quantity called the \gls*{QFI}, which can be obtained from a sum-rule integral of the dynamical susceptibility, and which certifies multipartite entanglement if its value exceeds classical expectations. Such an approach has recently led to the identification of delafossite KYbSe$_2$  as a proximate gapped $\mathbb{Z}_2$ \gls*{QSL} \cite{Scheie2024proximate}.

\gls*{RIXS} offers the exciting possibility of extending this type of quantum metrology to small samples and to non-equilibrium states (see Sec.~\ref{sec:non-equilibrium}) beyond the reach of \gls*{INS}. A key requirement to extend the established \gls*{QFI} methodology to \gls*{RIXS} is the need to extract the magnetic dynamical susceptibility in absolute units from \gls*{RIXS}. This will require the identification of a standard reference cross-section, such as the $dd$ excitation in cuprates, combined with theory to determine the appropriate normalization factor to map the \gls*{RIXS} operator to the spin operator \cite{Jia2016usig}. An alternative to converting \gls*{RIXS} to dynamical susceptibility would be to reformulate the \gls*{QFI} method in terms of the \gls*{RIXS}-operator. This approach, however, requires addressing the fact that the \gls*{RIXS}-operator in Eq.~\eqref{eq:KH_ME} is non-Hermitian \cite{Ren2024witnessing}. Such a theory would be very impactful by opening routes to using \gls*{RIXS} to measure the entanglement of other degrees of freedom such as orbital entanglement.

\subsection{Tunable QSLs}
A special area in \glspl*{QSL} research, and quantum materials more generally, is the study of quantum critical phenomena that occur when a magnetic phase transition takes place at absolute zero \cite{Paschen2021quantum}. A suite of novel approaches now enables the controlled modification of \glspl*{QSL} and access to such critical points; however, most require the use of small samples, which points to a special role for probing these systems with \gls*{RIXS}. An example of this would be few-layer exfoliated van der Waals materials. These may have fundamentally different microscopic properties from those in the bulk, as found in a recent study of monolayer $\alpha$-RuCl$_3$ \cite{Yang2023magnetic}. An especially intriguing possibility, in the same spirit, is to control the material dimensionality via thin-film deposition techniques. This includes growing \gls*{2D} perovskite phases in the in-plane direction to generate \gls*{1D} phases out of materials that form \gls*{2D} phases in the bulk \cite{Gruenewald2017engineering} or growing perovskite phases in the $[111]$ direction to create \gls*{2D} hexagonal motifs \cite{Lesne2023designing}.

\gls*{RIXS}'s access to small samples also facilitates studying systems subject to external tuning like applied strain. Such effects have already been studied theoretically with the proposal of accessing \glspl*{QSL} in spin ice materials \cite{Rochner2016spin}. Thin film or exfoliated geometries are also suitable for producing devices and applying gating voltages to tune the required interactions and realize \gls*{QSL} phases.

\section{Non-equilibrium phases}\label{sec:non-equilibrium}
\begin{figure*}[t]
\includegraphics[width=\textwidth]{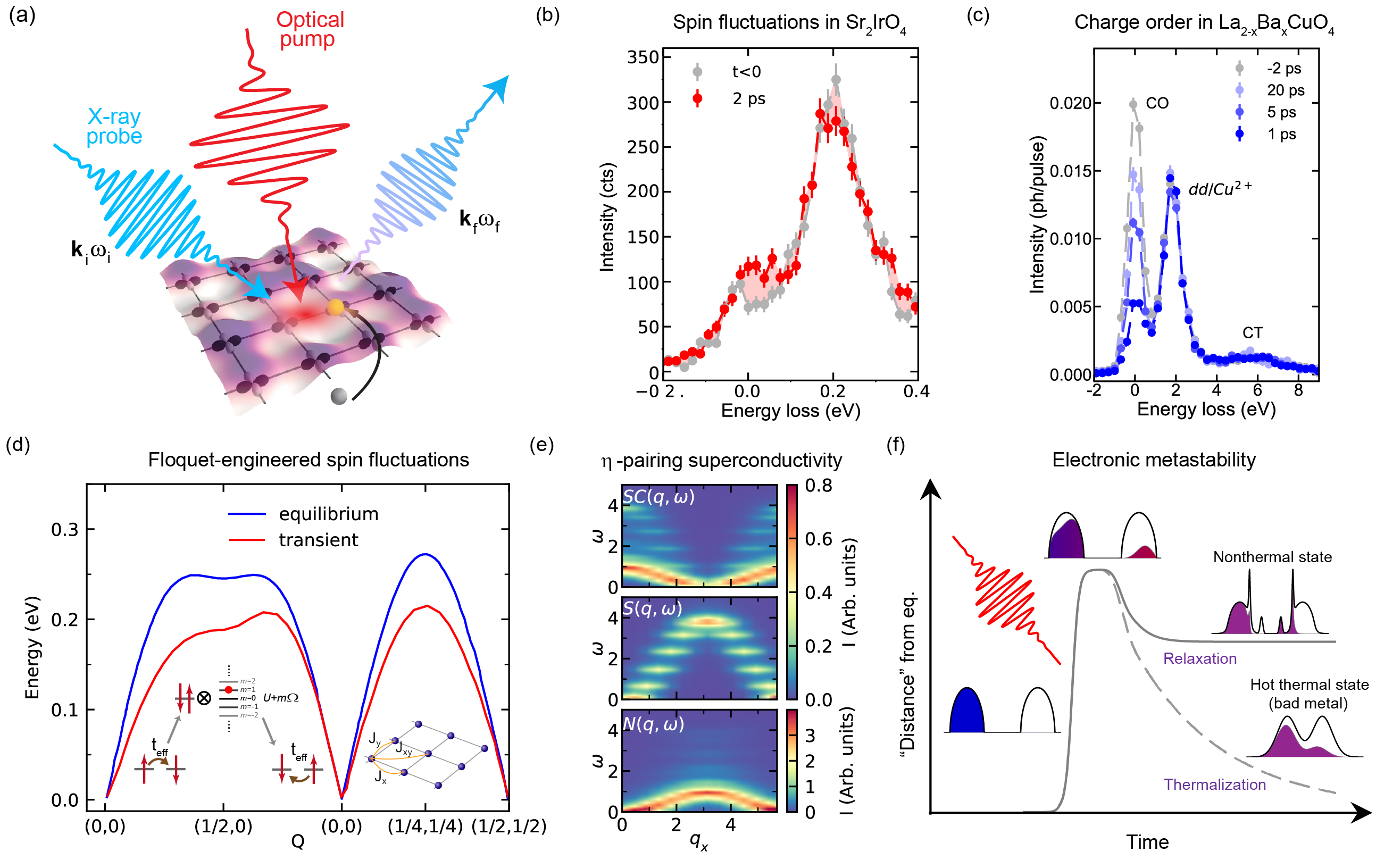}
\caption{\textbf{Time-resolved \gls*{RIXS} will enable the discovery of transient phases without equilibrium analogs.} (a) Sketch of a \gls*{trRIXS} experiment in which a material is perturbed by an optical pump and probed with short X-ray pulses \cite{Mitrano2020probing}. (b) Ir $L$-edge \gls*{trRIXS} spectra of pseudospin excitations of Sr$_2$IrO$_4$ after excitation with near-infrared photons  \cite{Dean2016ultrafast}. (c) Cu L$_3$-edge \gls*{trRIXS} spectra of photoexcited La$_{2-x}$Ba$_x$CuO$_4$ showing prompt melting of the charge order (CO) quasielastic peak \cite{Mitrano2019ultrafast}. (d)  Ultrafast pump fields can lift (lower) in energy the intermediate doubly-occupied state of the exchange process by an amount proportional to the number of virtually absorbed (emitted) pump photons and alter the spin wave dispersion across the entire \gls*{BZ} \cite{Wang2021xray}. $U$ is the onsite Coulomb repulsion, $t_{\textrm{eff}}$ the effective hopping amplitude, $\Omega$ the pump photon energy and $m$ the Floquet index \cite{Mitrano2020probing,Mitrano2020probing,Wang2021xray}. (e) Photoexcited low-dimensional Mott insulators are predicted to exhibit an emergent condensate called $\eta$-pairing, i.e., a staggered superconducting phase with characteristic charge [$N(q,\omega)$], spin [$S(q,\omega)$], and superconducting [$SC(q,\omega)$] response functions  \cite{Murakami2022exploring}. (f) Photodoped Mott insulators and quantum magnets can evade thermalization to a hot metallic state and give rise to nonthermal states (prethermal and metastable) if their dynamics is constrained by approximate conservation laws or symmetries \cite{Murakami2023photoinduced}.}
\label{fig_ultrafast}
\end{figure*}

Strong optical fields enable new protocols for controlling quantum materials and observing remarkable cooperative responses. Recent years have seen major successes in detecting emergent phenomena such as Floquet-engineered electronic band structures \cite{Wang2013observation, McIver2019light, Zhou2023pseudospin, Ito2023buildup} and new emergent magnetic \cite{Disa2020Polarizing, Afanasiev2021ultrafast, Disa2023photoinduced}, ferroelectric \cite{Kubacka2014large, Nova2019Metastable, Li2019terahertz}, topological \cite{Sie2018ultrafast}, and superconducting-like phases \cite{Fausti2011Light, Hu2014optically, Mitrano2016possible}. As we argue below, many aspects of these novel states are very difficult to probe with established techniques and are consequently not well understood. The emerging technique of \gls*{trRIXS} has particular strengths for probing transient phases and is poised to both improve our understanding of these dynamic states of matter and to discover new phenomena. These include strongly-correlated Floquet phases such as Floquet quantum magnets with dynamically-renormalized exchange interactions, broader forms of nonequilibrium condensation including $\eta$-pairing, and electronic metastability.

\subsection{The dawn of time-resolved RIXS}

Many photoinduced phenomena occur only on fast timescales of a few picoseconds or less, which tightly restricts how they can be measured. Optical spectroscopy is naturally well-suited to interrogate dynamics at femtosecond (or even attosecond) timescales and is a very popular ultrafast technique \cite{Giannetti2016ultrafast, Zhang2014dynamics, Torre2021colloquium}. There are, however, states that necessarily require accessing momentum-dependent information, including anti-ferro-type ordered or gapless states. Experimental tools capable of providing momentum resolution (see Box 1) are much more limited in this context, as many probes are intrinsically challenging to translate to fast timescales. The velocity ($\sim10^3$~ms$^{-1}$) and penetration depth ($\sim10^{-2}$~m) of thermal neutrons imply that a typical neutron scattering experiment will probe a material over a timescale of more than a microsecond, well beyond the lifetime of many light-induced states. Electrons represent a more appropriate tool for studying transient states, owing to their much shorter interaction length scale and the possibility of emitting them in short duration pulses. Ultrafast electron diffraction is an incisive probe of the transient state crystal structure \cite{Weathersby2015mega, Filippetto2022ultrafast}, while the photon-in, electron-out technique of time-resolved \gls*{ARPES} is a powerful probe of photo-excited band structures \cite{Bovensiepen2012elementary, Boschini2023time}. However, such experiments must be carefully designed to avoid so-called ``space charge'' effects, where the duration of the probe pulse gradually increases due to the mutual Coulomb repulsion between electrons in the bunch.

Being a photon-in, photon-out technique, \gls*{RIXS} is well-suited to probe matter at ultrafast timescales and in rather unique ways [see Fig.~\ref{fig_ultrafast}(a)]. The wavelength of x-ray photons is short enough to track light-induced dynamics across a substantial fraction (if not the entirety) of the \gls*{BZ}, well beyond the zero-momentum limit of ultrafast optical spectroscopy \cite{Giannetti2016ultrafast, Zhang2014dynamics, Torre2021colloquium}. Furthermore, owing to its rich cross-section, \gls*{trRIXS} is sensitive to the collective electronic excitation spectrum of quantum materials in the spin, charge, and orbital degrees of freedom. These electronic excitations with bosonic character are mostly invisible to time-resolved \gls*{ARPES} and to ultrafast x-ray and electron diffraction. Hence, \gls*{trRIXS} fills a long-standing gap in ultrafast experiments and constitutes the spectroscopy of choice for probing momentum-dependent, low-energy valence excitations in driven quantum materials. 

The field of \gls*{trRIXS} is currently at a watershed moment. These experiments require photon fluxes similar to or higher than those provided by synchrotrons, but in pulses of short duration and high intensity, which are typically generated by \gls*{XFEL} sources ~\cite{Emma2010first, McNeil2010xray, Bostedt2016Linac}. The first experiments tended to require implementing spectrometers at \glspl*{XFEL} specially for one-off pioneering experiments. Several experiments of this type have laid the foundation for the field. As shown in Fig.~\ref{fig_ultrafast}(b), antiferromagnetic Sr$_{n+1}$Ir$_n$O$_{3n+1}$ iridates \cite{Dean2016ultrafast, Cao2019ultrafast, Mazzone2021laser} (at the Ir $L$-edge) and CuGeO$_3$ spin chains \cite{Paris2021probing} (at the O $K$-edge) have provided the first snapshots of perturbed magnetic fluctuations at finite momentum and demonstrate that zero-moment optical excitations reshape the spin fluctuation spectrum throughout the entire \gls*{BZ}. In the cuprate La$_{2-x}$Ba$_x$CuO$_4$, \gls*{trRIXS} showed that the appearance of light-enhanced superconductivity is accompanied by a prompt suppression and sudden sliding of the charge order peak [see Fig.~\ref{fig_ultrafast}(c)], which recovers through diffusive dynamics involving the annihilation of topological defects \cite{Mitrano2019ultrafast, mitrano2019prb}. In the Mott-Hubbard system V$_2$O$_3$, a \gls*{trRIXS} study of the $dd$ excitations indicated the possible presence of a nonthermal state, where the excitation of electrons into the conduction band is likely accompanied by a modification of the V atom dimerization \cite{Parchenko2020orbital}. 

While promising, these first proof-of-principle experiments only explored a tiny fraction of the phase space opened by \gls*{trRIXS}. They did, however, motivate investments in dedicated  \gls*{trRIXS} facilities. The next generation of \gls*{trRIXS} experiments will benefit from technical advances, thanks to the simultaneous enhancement of spectral brightness through increased repetition rates (e.g.\ at the \gls*{LCLS}-II and the European XFEL) and the deployment of spectrometers with higher resolving power (of order 10,000) at \gls*{LCLS}-II, EXFEL, and SwissFEL. With these improvements in energy resolution and counting statistics, combined with more sophisticated and targeted photoexcitation schemes, \gls*{trRIXS} will be uniquely positioned to observe low-energy features of driven quantum phases at finite momentum and to gain qualitatively new insights into the microscopic physics of light-induced phenomena. The success of these experiments will depend on the development of accurate pump-probe synchronization schemes and detectors operating close to the XFEL repetition rate to resolve coherent dynamics in the time-dependent \gls*{RIXS} spectra. Further, it will be crucial to devise efficient energy dissipation schemes to mitigate average heating effects due to the increased repetition rate and tighter beam focusing.

\subsection{A unique view of emergent transient states}
A priority research direction for \gls*{trRIXS} is nonequilibrium superconductivity. In recent years, intense near- and mid-infrared fields have led to the observation of transient states featuring superconducting-like optical and magnetic properties \cite{Fausti2011Light, Hu2014optically, Mitrano2016possible, Suzuki2019photoinduced, fava2024magnetic}. These emergent electronic states defy the conventional expectation that external perturbations destroy quantum coherence by introducing heating and often exhibit novel symmetries \cite{Buzzi2020photomolecular} and energy scales \cite{Mitrano2016possible,Budden2021evidence}. 
The microscopic physics of these processes is still debated, with theoretical interpretations ranging from the melting of competing orders \cite{Raines2015enhancement}, to the amplification of the superconducting instability \cite{Babadi2017theory}, the cooling of phase fluctuations and low-energy quasiparticles \cite{Denny2015proposed, Nava2018cooling}, or light-induced pairing \cite{Kennes2017transient}. 

In copper oxides, light-induced superconductivity appears upon excitation with pulses tuned to the in-plane, bond-stretching \cite{Fausti2011Light}, and apical oxygen phonons \cite{Hu2014optically, Kaiser2014optically, Liu2020pump} or optical transitions across the charge-transfer gap \cite{Nicoletti2014optically}. Magnetic superexchange and strong oxygen-copper hybridization are vital ingredients for many theories of the equilibrium superconducting state of these materials, so they represent key elements for a microscopic theory of non-equilibrium superconductivity \cite{Scalapino2012common}.
\Gls*{trRIXS} can directly measure the magnetic superexchange by mapping the spin fluctuations spectrum at the Cu $L$-edge, while the charge-transfer energy is encoded in O $K$-edge spectra \cite{Shen2022role}. Both quantities are sensitive to the hopping matrix elements, which can be modulated either by atomic displacements due to nonlinear phonon coupling \cite{Forst2011nonlinear, Subedi2014theory, Disa2021engineering}, hybridization with strong electric fields \cite{Wang2021xray} [see Fig.~\ref{fig_ultrafast}(d)], or through a renormalization of the effective Coulomb repulsion \cite{Baykusheva2022ultrafast}. Future \gls*{trRIXS} measurements of near- and mid-infrared-driven cuprates such as YBa$_2$Cu$_3$O$_{6+\delta}$, supported by the theory described in Sec.~\ref{sec:trRIXS_theory}, will crucially contribute to the microscopic understanding of this intriguing phenomenon. \gls*{trRIXS} will also shed light on similarities and differences between light-induced superconductivity in the copper oxides and in FeSe$_{0.5}$Te$_{0.5}$ \cite{Isoyama2021light}, where Fe $L_3$-edge experiments \cite{Pelliciari2021evolution} could clarify the interplay between spin fluctuations and lattice degrees of freedom in the photoexcited state \cite{Gerber2017femtosecond}.

By further leveraging its spin sensitivity, \gls*{trRIXS} will directly access light-engineered magnetic interactions in a wider variety of quantum materials and dramatically expand the field of ultrafast magnetism \cite{Cao2019ultrafast, Mitrano2020probing, Torre2021colloquium}. Non-resonant pump pulses are expected to dress the electronic states of a magnetic system and renormalize its exchange energy scales, thus giving rise to a variety of Floquet magnetic phases [see Fig.~\ref{fig_ultrafast}(d)]. The transient exchange interaction can be enhanced or reduced depending on the pump energy, thus altering the finite-momentum spin excitations of the system \cite{Mentink2014ultrafast, Mentink2015ultrafast, chaudary2019orbital, Walldorf2019antiferromagnetic, Wang2021xray}. Alternatively, the spin collective modes can also be altered by distorting the lattice and tuning the inter-spin distance via nonlinear phonon coupling~\cite{Forst2011nonlinear, Subedi2014theory, Disa2021engineering}. High-resolution \gls*{trRIXS} at selected $L$-edges will map transient changes of the spin fluctuation spectrum in the driven state and experiments targeting a full reconstruction of the transient magnetic Hamiltonian are currently underway. Insulating oxides with high superexchange energies, such as one-dimensional (Sr$_2$CuO$_3$ \cite{Schlappa2012spin}, Sr$_{14}$Cu$_{24}$O$_{41}$ \cite{Schlappa2009collective}) and two-dimensional cuprates (La$_2$CuO$_4$ \cite{Braicovich2010magnetic}), nickelates (La$_{2-x}$Sr$_x$NiO$_4$ \cite{Fabbris2017doping}, and NdNiO$_2$ \cite{Lu2021magnetic}), and iridates (Sr$_2$IrO$_4$  \cite{Dean2016ultrafast}, and Sr$_3$Ir$_2$O$_7$  \cite{Mazzone2021laser}) represent a fertile ground for observing these effects. These compounds can be excited within the insulating gap and/or at resonance with high-energy phonons with minimal heating effects, while their magnetic excitation spectrum is accessible with a sub-100 meV energy resolution.

An intriguing target for future \gls*{trRIXS} studies is identifying novel forms of condensation.  Photoexcited Mott insulators are predicted to host an entirely new form of condensation, called $\eta$-pairing \cite{Yang1989eta, Zhang1990pseudospin, Kaneko2019photoinduced, Murakami2023spin, Murakami2022exploring,Murakami2023photoinduced}. The $\eta$-paired phase is a superfluid of doubly-occupied electronic states (doublons) with finite momentum \textbf{Q}$_{\eta}$ and associated with a broken SU(2) symmetry of the Hubbard Hamiltonian. This state is theoretically predicted to emerge in Mott insulators when photodoping holons and doublons by above-gap ultrafast excitation \cite{Kaneko2019photoinduced, Murakami2022exploring, Murakami2023spin} or by dynamically renormalizing the onsite Coulomb repulsion \cite{Peronaci2020enhancement}. The time and energy resolution of the current generation of \gls*{trRIXS} spectrometers is sufficient to test these theories in Cu $L_3$- and O $K$-edges measurements of undoped or doped Mott insulators, such as Sr$_2$CuO$_3$. If condensation and off-diagonal long-range order are established by the pump to make a state with $\eta$-pairing, this would induce clearly detectable signatures in the dynamical response. Such signatures include a divergent quasielastic pairing susceptibility at \textbf{Q}$_{\eta}$~\cite{Suzuki2018probing}, three collective modes at energies $\hbar\omega = 0,\:\pm(U - 2\mu)$ ($\mu$ being the chemical potential)~\cite{Zhang1990pseudospin}, and emergent replicas of collective modes in the structure factors [see Fig.~\ref{fig_ultrafast}(f)]. Crucially, this phase is gapless, and therefore challenging to detect with optical spectroscopy. The discovery of ${\eta}$-pairing would be groundbreaking as it would unveil a new type of nonequilibrium superconductivity in the absence of a parent equilibrium superconducting phase.

Finally, an important frontier of ultrafast research is represented by the quest to stabilize transient dynamical states and to search for metastable hidden phases [see Fig.~\ref{fig_ultrafast}(e)] that evade thermalization. While several metastable light-driven phases intimately involve structural rearrangements \cite{Nova2019Metastable,Stojchevska2014,Disa2023photoinduced}, purely electronic nonthermal states are rare. These states are called ``prethermal'' when their decay is prevented by approximate conservation laws or symmetry constraints, ``classically metastable'' when identifiable with a free energy minimum of emergent classical variables, or ``dissipative steady states'' when maintained out of equilibrium by a balance of driving and dissipation \cite{Murakami2023photoinduced}. 

\gls*{trRIXS} could track ultrafast pathways into new hidden/metastable states and help design these long-lived electronic states. Photoexcited Mott insulators and quantum magnets are promising systems for the observation of prethermal states activated by the selective melting of competing orders or by the realization of metastable electronic distributions. 
In the first case, \gls*{trRIXS} measurements of spin and $dd$ excitations \cite{Li2021unraveling} in photo-doped KCuF$_3$ could directly test the existence of a long-lived hidden state with dominant magnetic orders and flipped orbital pseudospin, which is theoretically predicted to occur in a driven three-quarter filled, two-band (d$_{y^2-z^2}$ and d$_{3x^2-r^2}$) Hubbard model \cite{Li2018theory}. 
Alternatively, tailored pump pulses could transiently break lattice symmetries and activate forbidden hopping pathways between structural units (e.g., a charge reservoir and a metallic plane) that are decoupled at equilibrium. Electrons would form a new quasi-steady-state distribution that is symmetry-protected against thermalization after photoexcitation. We anticipate this mechanism to be observable in low-dimensional, undoped and doped Mott insulators (e.g., Sr$_{14-x}$Ca$_x$Cu$_{24}$O$_{41}$), as well as in charge density wave systems (1T-TaS$_2$ \cite{Jia2023interplay}). \gls*{trRIXS} experiments at the transition metal $L$-edge or at the ligand $K$-edge can directly test for specific transfers of charge into the metastable state and will allow to accurately determine the nonequilibrium electronic distribution. The longer timescales of these phenomena will allow the use of more aggressive monochromatization schemes which, sustained by the high flux of high-repetition rate \glspl*{XFEL}, will enable experiments approaching a 10~meV energy resolution with adequate count rates.

\section{Functional materials}\label{sec:functional_materials}

\begin{figure}
\includegraphics[width=\columnwidth]{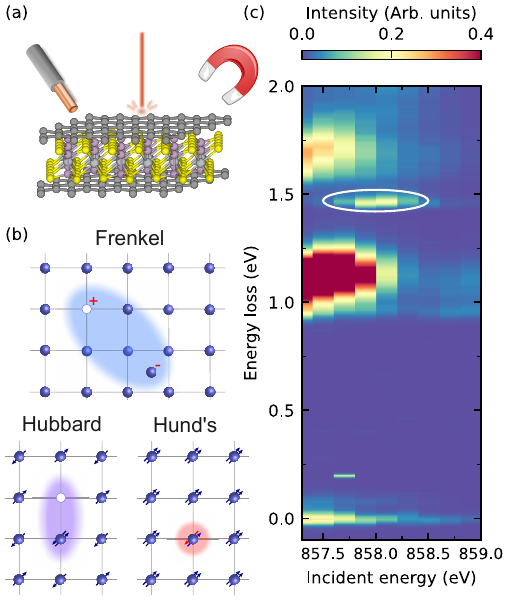}
\caption{\textbf{RIXS has unique assets for probing functional materials, such as the structure and propagation of novel magnetic excitons in van der Waals materials.} (a) van der Waals heterostructures interacting with light, electric, and magnetic fields, which offer routes to swtiching between different electronic phases and the transduction of information. (b) Illustration of Frenkel, Hubbard, and Hund's excitons. (c) \gls*{RIXS} is sensitive to nominally dipole-forbidden excitations and can detect excitons, such as the 1.45~eV feature (circled in white) in the Ni $L_3$-edge spectra of NiPS$_3$ \cite{Kang2020coherent}.}
\label{fig:excitons}
\end{figure}

As described in Sec.~\ref{sec:introduction}, quantum materials typically involve strong coupling between multiple charge, lattice, orbital, and spin degrees of freedom and high sensitivity to external perturbations such as electric and/or magnetic fields, strain, or photo-illumination. These properties offer promising routes towards new functionalities, which may provide the foundation for future technologies \cite{Tokura2017emergent}. For example, there are extensive efforts to devise methods to switch quantum materials between different electronic phases and/or to use quantum materials to transduce information between electrical, magnetic, or optical forms. Another major theme in this area is to incorporate quantum materials into functional devices. New developments in \gls*{RIXS} setups, coupled with the emergence of new platforms such as magnetic van der Waals materials, suggests that this nascent area may expand significantly in the coming years.

\subsection{Artificial materials and devices}
The ability of \gls*{RIXS} to deliver detailed spectroscopic information from thin layers using a small beam makes it well-suited for investigating functional quantum materials in device-like geometries. During the past decade \gls*{RIXS} has emerged as a powerful probe in these circumstances and has deepened our understanding of how magnetic and electronic phases can be tuned at interfaces in cuprate \cite{Dean2012spin}, iron \cite{Pelliciari2021evolution, Pelliciari2021tuning},  iridate \cite{Meyers2018decoupling}, nickelate \cite{Bisogni2016ground, Fabbris2016orbital}, and titanate-based \cite{Zhou2011localized} systems. 

Magnetic van der Waals materials has recently become a major focus of research \cite{Burch2018magnetism}. These materials are very diverse and often exhibit strong electron-electron interactions, and they are consequently being intensely studied as a platform for discovering novel correlated states. One particularly important aspect of these materials is their cleavability, which allows for the creation of heterostructures and device-like configurations. This property enhances our ability to control and investigate nanoscale phases under applied light, electric, and magnetic fields, as shown in Fig.~\ref{fig:excitons}(a). Extensive efforts are underway to realize new physics by stacking layers of different van der Waals materials, and leveraging charge transfer, strain, magnetic exchange,  and orbital hybridization as additional tuning knobs. However, electronic states at interfaces often change in poorly understood and uncontrolled ways. Thus far, only a handful of materials have exhibited stable excitons and magnetism down to the monolayer limit \cite{Gong2017discovery, Song2022evidence}. Understanding how electronic interactions are modified in these heterostructures is thus critical for successfully targeting new phenomena in this space. 

Although few-layer-thick exfoliated samples only intercept a small fraction of the x-ray beam, the feasibility of measuring these with \gls*{RIXS} has already been demonstrated \cite{Pelliciari2021evolution, DiScala2024elucidating}. For example, a recent study has indicated that the charge-transfer energy in NiPS$_3$ appears to be reduced in few-layer flakes compared to the bulk \cite{DiScala2024elucidating}. Here, the resonant nature of \gls*{RIXS} avoids background signals and allows one to selectively probe different layers within a heterostructure. This area will be bolstered by the delivery of nano-\gls*{RIXS} setups with x-ray beams of order 100~nm \footnote{One such beamline is the ARI beamline at the National Synchrotron Light Source II \url{https://www.bnl.gov/nsls2/beamlines/beamline.php?r=29-ID-2}} and will require parallel technical developments to mitigate new experimental challenges such as beam-heating and beam-damage. Preliminary work such as mounting samples directly on highly conductive surfaces (e.g., gold) and the use of capping layers made of inert materials has already demonstrated several potential routes to addressing these issues. Studying van der Waals devices \emph{in-situ} under applied electrical, strain and thermal fields is another frontier challenge. Van der Waals flakes are suitable for preparing pristine, small cross-section samples that can be used to achieve high field density. For the same reasons, these materials are also highly compatible with emerging methods for studying materials under high strain \cite{Qi2023recent}.

\subsection{Excitons}\label{sec:excitons}
Excitons are an emerging topic in \gls*{RIXS} studies of functional materials. These collective excitations play a crucial role in determining the optical properties of quantum materials and have numerous applications in photonics and energy science \cite{Wang2018colloquium}. Excitons are more likely to form in low-dimensional materials, where the reduced screening of the Coulomb interactions promotes their formation. Hence, van der Waals materials, in either bulk or exfoliated form, are one of the most appropriate platforms for realizing and studying them. One of the most common types of exciton are the Frenkel-type, which usually form in weakly correlated semiconductors and are composed of electron-hole pairs bound by the screened Coulomb interaction \cite{Frenkel1931on, Yu2010fundamentals} [see Fig.~\ref{fig:excitons}(b)]. In most cases, Frenkel excitons are well understood using approaches based on the Bethe-Salpeter equation \cite{Louie2021discovering}. Excitons in strongly correlated quantum materials are less understood and represent an active research area. In strongly correlated systems, local Coulomb and Hund's exchange interactions play a major role in exciton formation and the exciton can no longer be accurately modeled in terms of the bound states of two interacting single-particle wavefunctions. As seen in Fig.~\ref{fig:excitons}(b), Hubbard and Hund's excitons modify the local spin configuration within the sample, which means that these excitons couple strongly to magnetism. Although many-body excitons exist in several classes of compounds, van der Waals materials have been shown to host unique coupling between excitons and magnetism, including magnetic state-dependent polarized light emission \cite{Seyler2018ligand, Dirnberger2022spin} and exciton energy \cite{Bae2022exciton}. These phenomena offer compelling possibilities for new types of magneto-optical physics and functionalities \cite{Gong2019two} and so far appear to be unique to van der Waals materials. 

A fundamental challenge in studying strongly correlated excitons is that many of them are nominally optical dipole forbidden and the theoretical description of their optical cross-section is the subject of active research \cite{Louie2021discovering}. Without an established theory for the optical response of correlated excitons, it can be challenging to definitively determine their electronic structure.  Since \gls*{RIXS} involves two dipole transitions, excitons can be directly excited in the scattering process [see Fig.~\ref{fig:excitons}(c)] as described by the \gls*{KH} equation. Thus, \gls*{RIXS}, combined with theory as described in Sec.~\ref{sec:Theory}, provides a powerful means of determining the electronic character of these excitons. 

Another special asset of \gls*{RIXS} is its ability to probe the exciton's momentum-dependent dispersion, which will allow high-energy resolution studies to determine how magnetic van der Waals excitons propagate, as has already been done via Ni $L$-edge measurements of NiPS$_3$, NiCl$_2$, NiBr$_2$, and NiI$_2$ \cite{He2024magnetically, Occhialini2024nature}. Given that these excitons behave as spin vacancies within the magnetic lattice, it is likely that propagating excitons will be dressed by magnetic excitations. Measurements of exciton dispersion offer the possibility of extracting the detailed exciton-magnon interaction, which underlies modern proposals for exciton-based coherent magnon sensing \cite{Bae2022exciton}. A better understanding of magnetic exchange bias in heterostructures that couple exciton-hosting and magnetically ordered layers offers pathways to control the chirality of exciton emission \cite{Seyler2018valley, Zhong2017van}.

Upon interaction with light, excitonic states hybridize with the electromagnetic field, giving rise to emergent exciton-polariton excitations involving collective oscillations of the polarization in the solid. These states have been recently identified in breakthrough experiments in illuminated magnetic van der Waals materials such as NiPS$_3$ and CrSBr \cite{Dirnberger2022spin, Wang2023magnetically, Dirnberger2023magneto}. These types of exciton are being heavily investigated in the context of the detection, harvesting, emission, propagation, and modulation of light \cite{Basov2016polariton} and coupling \gls*{RIXS} with continuous sample illumination will enable new approaches to studying the spectroscopic properties of exciton polaritons at previously inaccessible momenta.

\section{Theory}\label{sec:Theory}

\begin{figure*}
\includegraphics[width=\textwidth]{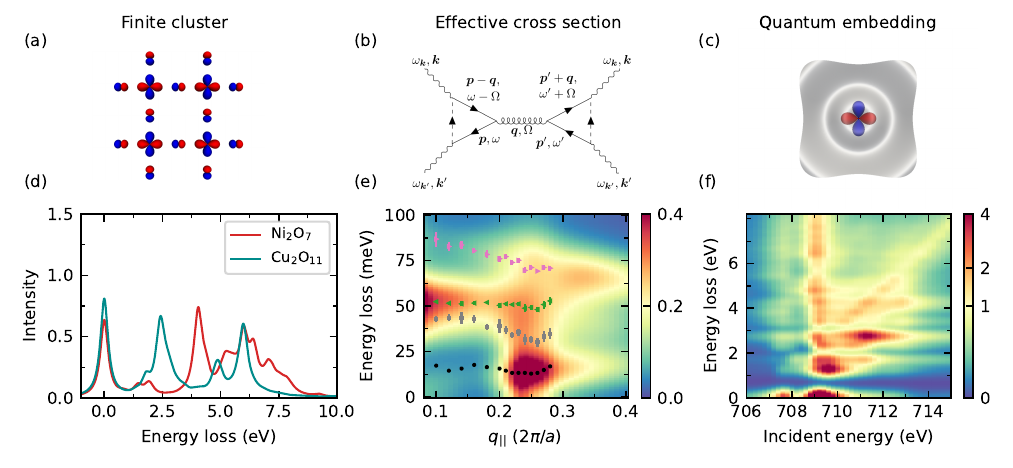}
\caption{\textbf{Summary of approaches to calculating \gls*{RIXS}: finite cluster, effective cross-section, quantum embedding.} (a) Finite cluster methods, where the \gls*{RIXS} cross-section is evaluated using the exact many-body eigenstates of a small cluster; (b) effective models, where the \gls*{RIXS} cross-section is replaced with a suitable approximation; and (c) quantum embedding methods, where the continuum of final-state excitations is captured using a dynamical mean-field. Panel (d) shows oxygen $K$-edge \gls*{RIXS} spectra obtained from small cluster models for the cuprate La$_{1.65}$Sr$_{0.35}$CuO$_4$ and low-valence nickelate La$_4$Ni$_3$O$_8$ materials~\cite{Shen2022role}. Panel (e) shows the model phonon excitations La$_{2-x}$Sr$_x$CuO$_4$ compared against the energies of the modes measured at the O $K$-edge~\cite{Huang2021quantum}. Panel (f) shows dynamical mean field theory (DMFT)-predicted Fe $L$-edge spectra for Fe$_2$O$_3$ with a mixture of fluorescence-like and Raman-like excitations~\cite{Hariki2020LDA}. 
\label{fig:scales}}
\end{figure*}

As emphasized in Sec.~\ref{sec:cross_section}, \gls*{RIXS} spectra are information-rich, but some of this information is only accessible through a detailed understanding of the cross-section, including the core hole's influence on the system. The field has, consequently, benefited significantly from close collaborations between theory and experiment. In this section, we discuss some of the major theoretical approaches for modeling \gls*{RIXS} experiments, emphasizing what we believe are the remaining challenges and how the field will evolve to tackle them in the coming years. 

\subsection{Finite Clusters}
Small cluster methods are a direct approach for treating the complexity of the \gls*{RIXS} cross-section. This class of methods approximates the infinite system with a finite cluster that is small enough that it can be solved exactly using a suitable non-perturbative numerical method, as illustrated in Fig.~\ref{fig:scales}(a). The \gls*{KH} cross-section is then evaluated directly using the exact many-body eigenstates of the cluster while accounting for the system's many-body interactions and the influence of the core hole. Due to the finite extent of the cluster, these methods are appropriate for describing experiments where the length scales of the targeted excitations are comparable to the cluster size. For example, small cluster \gls*{ED} played a pivotal role in understanding local $dd$ \cite{Ghiringhelli2009, Chen2010unraveling, Moretti2011energy} and other orbital/charge excitations \cite{Okada2002copper, Okada2003theory, Tsutsui2003mott, Okada2004effects, Vernay2008Cu, Chen2010unraveling, Shen2022role} in strongly correlated systems [see Fig.~\ref{fig:scales}(d)]. Cluster-based approaches have also been applied to many-body exciton states, as described in Sec.~\ref{sec:excitons}, provided that the spatial extent of the exciton is small enough to be captured within the cluster \cite{Kang2020coherent, He2024magnetically, Occhialini2024nature}. Nowadays, codes for performing these kinds of calculations are quite advanced~\cite{Quanty, Wang2019EDRIXS, EDRIXS} and we envisage that the research landscape will pivot to incorporating these tools into standard experimental workflows, for example by using machine learning to automatically extract model Hamiltonians from the data and to incorporate real-time theory feedback to optimize data collection during beamtimes \cite{Johnston2022perspective, Chen2021machine, Samarakoon2021machine}.

Modeling the excitations of an interacting quantum system requires knowledge of its many-body wavefunctions. Because the size of the associated Fock space increases factorially in particle number and system size, cluster sizes are generally quite small, resulting in large energy spacing and coarse momentum resolution. For this reason, cluster methods often struggle to capture systems with a continuum of final-state excitations~\cite{Ishii2005momentum} (with the notable exception of quasi-1D systems~\cite{KourtisPRB2012, Wohlfeld2013microscopic, Schlappa2018probing, Kumar2021Unraveling, Kumar2019Ladders}) or systems with strong \gls*{EPC}~\cite{Ament2011determining, Lee2013role, Johnston2016electron, Geondzhian2020generalization} (see Sec.~\ref{sec:th_EPC}). Thus, there is an urgent need to develop more advanced cluster solvers. The future of this topic is likely to focus on co-opting and extending existing advanced many-body numerical methods to calculate \gls*{RIXS}. Recent progress in this direction includes the introduction of \gls*{DMRG}-based solvers \cite{Nocera2018computing, zawadzki2023timedependent} for quasi-1D systems. We expect that these approaches will play an increasingly important role in the future, as modern \gls*{DMRG} implementations continue to push into the second dimension~\cite{Jiang2021groundstate, Potapova2023dimensional}. 

Another promising approach to reducing the computational complexity of cluster methods is to explicitly model the photon absorption and emission events in the time domain~\cite{chen2019theory, Lai2019ultrafast, Werner2021nonequilibriumb, zawadzki2023timedependent}. This approach reduces the calculation of the \gls*{RIXS} cross-section to time evolving the system during the scattering event, which can be done efficiently using, for example, time-dependent \gls*{DMRG} in the case of a \gls*{1D} system. Here, the optical absorption process can be treated classically to further reduce the computational complexity of the problem~\cite{zawadzki2023timedependent}. This approach can also be easily extended to non-equilibrium settings (see Sec.~\ref{sec:trRIXS_theory}), and will thus play a prominent role in the interpretation of the experiments outlined in Sec.~\ref{sec:non-equilibrium}. 

\subsection{Effective cross-sections}
While one can express the \gls*{RIXS} intensity as the Fourier transform of a two-particle correlation function, the relevant operators $O_{\bf q}(t)$ are complicated and challenging to identify. Their exact form depends on the material being studied, the x-ray edge, and the energy range probed by the photons. Nevertheless, procedures have been developed for systematically formulating effective theories of \gls*{RIXS} in terms of ``simpler'' correlation functions that can be easier to compute and interpret [see Fig.~\ref{fig:scales}(b)]. 

An early and highly influential approach along these lines is the \gls*{UCL} expansion. It exploits the fact that the core hole lifetime is often the shortest time scale in the problem~\cite{Brink2005correlation}, allowing one to expand the \gls*{KH} cross-section as a series of multi-particle correlation functions. The lowest-order terms ($\propto 2/\Gamma$) are closely related to the pure charge, spin, and orbital response functions, while higher-order terms involve multi-particle correlation functions describing coupled spin/charge/orbital excitations~\cite{Brink2005correlation, Ament2007ultrashort, Jia2016usig}. The \gls*{UCL} has been widely applied to various materials but has been most successful in describing low-energy collective excitations in materials whose energy is a small fraction of the inverse core hole lifetime $\Gamma/2$. For example, it played a key role in early studies of magnetic excitations in high-$T_\mathrm{c}$ cuprates~\cite{Braicovich2010magnetic, Bisogni2012Bimagnon}. 

Effective theories like the \gls*{UCL} provide a direct means to compute \gls*{RIXS} spectra using established many-body methods. For example, while the multi-particle correlation functions may be complex, they can be calculated using methods like \gls*{QMC} or \gls*{DMRG}. So far, studies like these have focused on $L$-edge measurements of systems with nearly complete valence shells (e.g., $3d^8$ or $3d^9$). For example, the low-energy magnetic excitations of the \gls*{AFM} spin-1 chain Y$_2$BaNiO$_5$ probed at the Ni $L$-edge have been described using a combination of dipolar and quadrupolar spin correlation functions computed using \gls*{DMRG}~\cite{Nag2022quadrupolar}, while \gls*{QMC} calculations of the dynamical spin structure factor have been used as proxies for the Cu $L$-edge spectra of 1D~\cite{Kumar2019Ladders, Kumar2021Unraveling} and 2D cuprates~\cite{Jia2014persistent}. However, this approach will face key challenges moving forward. Low-order expansions can be slow to converge to the full \gls*{KH} intensity in a way that depends on the scattering geometry~\cite{Jia2014persistent}. Therefore, the expansion order will vary for different edges and scattering conditions. A low-order expansion has only been derived for $L$-edge measurements on $d^9$ systems~\cite{Jia2014persistent} and generalizations to new systems or different edges will need to be worked out on a case-by-case basis. In this context, an alternative operator expansion formalism~\cite{Lu2017nonperturbative}, which expands the \gls*{RIXS} operator in terms of local operators acting in the vicinity of the core-hole site, could provide a more direct path forward. This approach directly circumvents the need for a small expansion parameter and dispenses with the requirement of making assumptions about the valence state of the system.

Feynman diagram methods are another emerging approach to formulating effective \gls*{RIXS} theories. Here, the cross-section is cast as perturbation series of contributions to the \gls*{RIXS} amplitude, as represented by a series of Feynman diagrams, like the one-phonon diagram shown in Fig.~\ref{fig:scales}(b)~\cite{Devereaux2016directly}. Here, the relevant diagrams are evaluated using suitable propagators. For example, lattice excitations in high-$T_\mathrm{c}$ cuprates probed at the Cu $L$- and O $K$-edges~\cite{Devereaux2016directly, Huang2021quantum} [see Fig.~\ref{fig:scales}(e)] have been modeled with very high momentum resolution using non-interacting electron and phonon propagators. As with the \gls*{UCL} expansions, this approach can be systematically improved by using {\it dressed} electron and phonon propagators or by adding higher-order diagrams. For example, collective spin and charge excitations can be described by replacing the phonon propagator in Fig.~\ref{fig:scales}(b) with the appropriate dynamical susceptibility~\cite{Tsvelik2019resonant}. Such an approach has already been used to model O $K$-edge measurements on high-$T_c$ cuprates, where the relevant propagators were derived from cluster perturbation theory~\cite{Matsubayashi2023numerical}. 

\subsection{Quantum embedding}
While \gls*{RIXS} studies of correlated metals and semimetals are becoming increasing prevalent, many of the finite cluster approaches described above lack the energy and momentum resolution need to accurately describe their continuum of final states and fluorescence features~\cite{Ghiringhelli2005NiO, Bisogni2016ground}. 
A current research frontier involves developing quantum embedding methods like \gls*{DMFT} to capture this aspect of the data [see Fig.~\ref{fig:scales}(c)]~\cite{Bisogni2016ground, Hariki2020LDA, Winder2020xray, Higashi2021core}. 
Here, the impurity site captures the local correlations and multiplet effects, while the final state continuum is encoded in the dynamical mean-field. These calculations can thus describe both Raman- and fluorescence-like excitations in experiments such as that shown in Fig.~\ref{fig:scales}(f). However, a critical and unresolved question relates to how one should compare the predicted spectra to experiments. \gls*{RIXS} is a coherent scattering process, which sums over all possible core-hole sites in the lattice [see Eq.~\eqref{eq:KH_ME}]. Conversely, \gls*{DMFT} methods based on Anderson impurity models only simulate the core hole at the impurity site while leaving the mean-field bath unchanged in the intermediate state. The calculated \gls*{RIXS} response is thus akin to an unmeasurable ``local'' response, where the core hole is only created at a single site. While one can leverage cluster extensions like \gls*{DCA} to re-encode momentum into the problem, wave-function-based cluster solvers are required to directly evaluate Eqs.~\eqref{eq:KH} \& \eqref{eq:KH_ME}. \gls*{ED} would limit the cluster sizes and number of bath levels and thus severely limit momentum and energy resolution. Alternatively, spectra could be formulated using the aforementioned operator or \gls*{UCL} expansions, where the corresponding correlation functions would be computed using an embedded cluster framework. 

Cumulant expansions of the Green's function also show promise for calculating the \gls*{RIXS} response of correlated metals~\cite{Gilmore2021description}. This method also offers a key advantage of being easily coupled with {\it ab initio} methods, allowing one to fully account for the crystal structure's chemical complexity and periodicity. The recent progress in developing accurate correlation-exchange functionals for strongly correlated systems~\cite{Sun2015strongly, Sun2016accurate} also means that these {\it ab initio}-based techniques could be more broadly applied to strongly correlated materials in the future; however, detailed benchmarking against nonperturbative approaches like small cluster \gls*{ED} or \gls*{DMRG} will be necessary.

\subsection{Electron-phonon interactions}\label{sec:th_EPC}
Another area where theory has been out paced by experiments is in understanding how \gls*{EPC} can be probed using \gls*{RIXS}. With improved instrument resolution, experiments now frequently resolve low-energy phonon excitations~\cite{Yavas2010observation, Lee2013role, Johnston2016electron, Moser2015electron, Fatale2016hybridization, Chaix2017dispersive, Meyers2018decoupling, Rossi2019experimental, Peng2020enhanced, Braicovich2020determining, Dashwood2021probing}. 
This development is significant because early theoretical modeling suggests that the $\bm{q}$ dependence of the \gls*{EPC} coupling constant $g(\bm{k},\bm{q})$ can be extracted from the intensities of the phonon excitations~\cite{Ament2011determining}. Subsequently, several groups have attempted to map the \gls*{EPC} in materials such as the cuprates~\cite{Chaix2017dispersive, Braicovich2020determining, Lin2020strongly, Peng2020enhanced, Rossi2019experimental}, where the exact role of this interaction remains controversial. Accessing lattice excitations with \gls*{RIXS} also provides new opportunities to examine their interplay with spin, orbital, and charge excitations. Indeed, Franck-Condon-like dressing of local $dd$ excitations~\cite{Hancock2010lattice, Braicovich2010magnetic, Yavas2010observation, Lee2015charge} and \gls*{EPC}-induced renormalizations of magnetic~\cite{Lee2013role} and electronic~\cite{Johnston2016electron} energy scales have been inferred from such experiments. These experiments can also provide access to orbital-resolved \gls*{EPC} \cite{Lee2013role} and couplings far from the Fermi level~\cite{Dashwood2021probing}, which are extremely difficult to extract from other experiments. 

Modeling lattice excitations in \gls*{RIXS} experiments is extremely challenging and remains a crucial problem for the community. For this reason, most quantitative analysis has focused on Holstein-like interactions and relied on single-site models that neglect electron mobility ~\cite{Ament2011determining, Geondzhian2020generalization}. While these assumptions result in closed-form expressions for the phonon excitations, they drastically oversimplify the problem. More recent studies have found that relaxing these approximations can qualitatively change the results. For example, local coupling to multiple phonon modes~\cite{Geondzhian2020generalization}, changes in local harmonic potentials in the intermediate state~\cite{Geondzhian2020generalization}, reintroducing electron mobility~\cite{Bieniasz2021beyond}, and dispersive phonons~\cite{Bieniasz2022theory} all impact the predicted spectra. It is, therefore, highly desirable to develop more robust methods for predicting lattice excitations in \gls*{RIXS} experiments. In this context, the most viable option is likely to be combining diagrammatic approaches~\cite{Devereaux2016directly, Tsvelik2019resonant} with many-body methods for efficiently computing electron and phonon Green's functions of \gls*{EPC} Hamiltonians. Modeling of nonlinear \gls*{EPC} and lattice anharmonicity will also be needed to guide future non-equilibirum experiments leveraging nonlinear phononics, where the atoms are significantly displaced from their equilibrium positions. 

Finally, some have argued that the lattice responds to the combined charge distribution of the core hole and excited electron in the intermediate state~\cite{Geondzhian2018demonstration, Gilmore2023quantifying}. Thus, the \gls*{EPC} probed by \gls*{RIXS} may in reality be an exciton-lattice coupling, with important implications for interpreting and analyzing phonon excitations. One theoretical analysis has shown that the lattice-core-hole coupling can frustrate polaronic effects on the site where the core hole is created, thus enhancing the effects of electron mobility on the spectra~\cite{Bieniasz2021beyond}. To address this idea, rigorous calculations for the strength of the core-hole-lattice coupling are needed, together with frameworks capable of capturing itinerancy and other effects. In this context, systematic comparisons between theory and experiment on either quasi-1D (e.g., Sr$_2$CuO$_3$ \cite{Schlappa2018probing} or Li$_2$CuO$_2$ \cite{Johnston2016electron}) or quasi-0D materials (e.g., CuB$_2$O$_4$ \cite{Hancock2010lattice}), where the spectra can be reliably computed with existing \gls*{DMRG} or \gls*{ED} approaches would be extremely valuable. 

\subsection{Time-resolved RIXS theory}\label{sec:trRIXS_theory}
Another emerging and critical area for theory is the development of methods for describing ultrafast pump-probe \gls*{RIXS} experiments (see also Sec.~\ref{sec:non-equilibrium})~\cite{Wang2018theoretical, Cao2019ultrafast, Mitrano2020probing}. Early breakthrough work has been done in deriving a theory of \gls*{trRIXS} using the Keldysh formalism~\cite{chen2019theory}. This approach captures the full dynamics of the problem, including any overlap between the pump and the core-hole excitation, resulting in a four-time correlation function that is numerically expensive to compute. Applications for strongly correlated systems have thus been restricted to small Hubbard clusters and very short timescales~\cite{Wang2020time}. As such, this approach is well-suited to processes such as Floquet effects that occur while the laser pulse is interacting with a material \cite{Wang2021xray}.

Developing alternative approaches to computing \gls*{trRIXS} that predict the dynamical evolution after pump excitation will be important for future progress. A very promising approach is to explicitly treat the photon absorption and emission events in the scattering process in a time-dependent framework~\cite{Werner2021nonequilibrium, zawadzki2023timedependent}, which can readily be extended to non-equilibrium \gls*{RIXS} experiments with reduced computational costs. This approach is expected to facilitate simulations that address longer timescales and is already being applied in \gls*{DMFT}-based methodologies~\cite{Werner2021nonequilibriumb, Eckstein2021simulation}, where they are providing insights into nonequilibirum charge dynamics~\cite{Werner2021nonequilibriumb} and lattice excitations~\cite{Werner2021nonequilibrium} in correlated systems. These approaches are particularly useful for using \gls*{RIXS} to detect dynamical changes in the effective Coulomb repulsion after photo-excitation~\cite{Tancogne-Dejean2018ultrafast,Baykusheva2022ultrafast}.

Another important theoretical advance would be a formal systematic derivation of a non-equilibrium version of the \gls*{UCL} or operator expansions for the \gls*{RIXS} cross-section. Using such an approach, it may be possible to reduce the four-time correlation function to a series of effective two-time correlation functions that will be easier to compute. While this approach could miss important interference effects when the core-hole lifetime and pump overlap, it may enable calculations to address slow dynamics or situations where energy transfer between different degrees of freedom is crucial.

\section{The future is bright}\label{sec:Outlook}

\gls*{RIXS} has established itself as a powerful probe of quantum materials, and is set to make key contributions to areas ranging from strongly correlated materials (Sec.~\ref{sec:strange}), to spin liquids (Sec.~\ref{sec:QSL}), to non-equilibrium (Sec.~\ref{sec:non-equilibrium}) and functional (Sec.~\ref{sec:functional_materials}) phases. \gls*{RIXS} is also increasingly influential in other topics in quantum materials research not covered here, including heavy fermions or Kondo lattices, as well as areas such as catalysis and batteries. This diversity is all the more remarkable in light of the relative youth of this technique. For example, the first measurements of magnons date to only about fifteen years ago. Now, \gls*{RIXS} is not only measuring magnons, but doing so under extreme conditions of atomically thin layers and ultrafast transient states. The growing maturity of \gls*{RIXS} is also becoming more apparent. For example, many of the complexities of the \gls*{RIXS} cross-section are now being exploited to gain new physical insights, rather than being regarded as a challenge for interpreting spectra.

To date, \gls*{RIXS} has proven particularly impactful in studying correlated insulators and their derived phases like strange metals. This is largely because correlated insulators tend to exhibit comparatively simpler, more easily interpretable spectra. One new direction for \gls*{RIXS} research will focus on extending its application to weakly or moderately correlated topological metals. For instance, \gls*{RIXS} could significantly contribute to the emerging field of Kagome metals, which exhibit unique flat band, topological, and charge-order behaviors. Notably, some of these materials feature multiple active magnetic ions, which presents opportunities for \gls*{RIXS} to disentangle modes arising from different sub-lattices. Another promising avenue is the recent identification of a new class of altermagnetic materials, which hold potential for novel applications. Polarization-dependent \gls*{RIXS} might be particularly effective in identifying spin-based magnon band splitting, which is crucial for understanding the novel magneto-transport behaviors in altermagnets.

Looking further into the future, it is clear that major innovation often arises at the interface between different fields. Entanglement plays a pivotal role in quantum materials, making it highly desirable to develop new methods for its manipulation. A frontier challenge lies at the intersection of entanglement metrology and ultrafast techniques: can we use lasers to transiently enhance quantum entanglement? Tackling the complexities of both fields simultaneously presents a significant challenge, but the potential rewards are compelling. Achieving success in this endeavor could position \gls*{RIXS} as a unique tool to identify novel transient quantum phases. Quantum phase transitions, occurring when a low-temperature ordered phase is suppressed via a non-thermal tuning parameter, have long intrigued the physics community. This fascination stems from the potential for such transitions to create highly entangled phases, potentially leading to new and exotic states. Can similar phenomena can be induced through ultrafast photoexcitation? While photoexcitation can certainly modulate order parameters, a key goal of ultrafast science is to develop optimal control schemes for the on-demand manipulation of light-induced entangled states, such as light-driven superconductors.

The combination of \gls*{RIXS} with applied fields also represents an expanding area within the field of functional materials. A major goal here is to study how materials can be switched to different phases in situ. In the future, these in-situ studies could be expanded to measure the energy gain side of the \gls*{RIXS} spectrum under electric or thermal gradients. This ambitious idea aims to directly identify and study which collective excitations are propagating in response to these field gradients. This type of information would help us understand the processes relevant to spintronics in unprecedented detail. 

It should be emphasized that the technology behind \gls*{RIXS} and time-resolved \gls*{RIXS} is still advancing rapidly, pushing the energy resolution towards and below the 10~meV mark, and each improvement in the energy resolution is bound to expand the applicability of \gls*{RIXS} to new materials and to reveal new excitations and physical phenomena. Sub-10~meV has already been demonstrated in a proof-of-principle experiment in the hard x-ray regime, using novel quartz-based flat-crystal optics \cite{Kim2018quartz} and spectrometers targeting sub-10~meV in the soft x-ray regime are currently being commissioned \cite{Miyawaki2022design}. With several orders of magnitude higher photon flux available at new \gls*{XFEL} sources, implementation of such novel spectrometer schemes for quantum materials research is feasible, and we can expect clearer and higher precision access to low-energy quasiparticles and fast transient timescales going forward. The continued refinement of existing technologies combined with the development of ambitious ideas for improved resolution such as \gls*{XFEL} oscillators, critical-angle transmission gratings, and asymmetrically cut Bragg optics \cite{Margraf2023low, Shvydko2020diffraction, Heilmann2022xray}, means that \gls*{RIXS} is set to continue to shine light on many of the most fascinating problems in quantum materials.

\begin{acknowledgments}
The authors would like to thank all those who have collaborated with us in pushing this research field forward over the years. We acknowledge Ji{\v r}{\' i} Chaloupka, G{\'a}bor Hal{\'a}sz, Atsushi Hariki, Di-jing Huang, Jungho Kim, Wei-Sheng Lee, Jan Kune{\v s}, Yao Shen, Justina Schlappa, Thorsten Schmitt, Philip Werner, and Yao Wang for their willingness to share data and we thank Edoardo Baldini, Valentina Bisogni, Tom Devereaux, Joe Dvorak, Martin Eckstein, Paul Evans, Wei-Sheng Lee, Chi-Chang Kao, Daniel Mazzone, and Kejin Zhou for comments on the manuscript. Work by M.~M.\ and M.~P.~M.~D.\ was supported by the U.S.\ Department of Energy (DOE), Division of Materials Science, under Contract No.~DE-SC0012704. Work by S.~J. was supported by the National Science Foundation under Grant No.~DMR-1842056. Work by Y.-J.~K.\ was supported by the Natural Science and Engineering Research Council (NSERC) of Canada. 
\end{acknowledgments}

\bibliography{refs}

\clearpage

\end{document}